\documentclass{amsproc}
\usepackage{graphicx,mathrsfs}
\usepackage{tikz}
\usepackage{booktabs}
\usepackage{tipa}

\theoremstyle{definition}

\theoremstyle{remark}

\numberwithin{equation}{section}

\newcommand{\T}{\mathcal{T}}

\def\Tr{\text{Tr}}

\newcommand{\mapsfrom}{\mathrel{\reflectbox{\ensuremath{\mapsto}}}}

\def\m@th{\mathsurround=0pt}
\mathchardef\bracell="0365
\def\upbrall{$\m@th\bracell$}
\def\undertilde#1{\mathop{\vtop{\ialign{##\crcr
   $\hfil\displaystyle{#1}\hfil$\crcr
    \noalign
    {\kern1.5pt\nointerlineskip}
    \upbrall\crcr\noalign{\kern1pt
  }}}}\limits}

\mathchardef\hatbracell="0362
\def\hatupbrall{$\m@th\hatbracell$}
\def\underhat#1{\mathop{\vtop{\ialign{##\crcr
   $\hfil\displaystyle{#1}\hfil$\crcr
    \noalign
    {\kern1.5pt\nointerlineskip} 
    \hatupbrall\crcr\noalign{\kern1pt 
  }}}}\limits}

\begin{document}

\title[Twisted reductions of integrable lattice eqs. and their Lax reps.]{Twisted reductions of integrable lattice equations,\\
and their Lax representations}

\author[C.M.Ormerod]{Christopher M. Ormerod$^1$}
\author[P. H. van der Kamp]{Peter H. van der Kamp$^2$}
\author[J. Hietarinta]{Jarmo Hietarinta$^3$}
\author[G. R. W. Quispel]{G. R. W. Quispel.$^2$\\[10mm]
{\small
$^1$ Department of Mathematics, California Institute of Technology, Pasadena, CA, 91125, USA,\\
$^2$ Department of Mathematics and Statistics, La Trobe University, 3086, VIC, Australia,\\
$^3$ Department of Physics and Astronomy, University of Turku, 20014 Turku, Finland.}}

\email{christopher.ormerod@gmail.com}

\begin{abstract}
It is well known that from two-dimensional lattice equations one can
derive one-dimensional lattice equations by imposing periodicity in
some direction. In this paper we generalize the periodicity condition
by adding a symmetry transformation and apply
this idea to autonomous and non-autonomous
lattice equations. As results of this approach, we obtain new
reductions of the discrete potential Korteweg-de Vries equation,
discrete modified Korteweg-de Vries equation and the discrete
Schwarzian Korteweg-de Vries equation. We will also describe a direct
method for obtaining Lax representations for the reduced equations.
\end{abstract}

\maketitle

\section{Introduction}
A key property of integrable partial differential equations (PDEs) is
the existence of multisoliton solutions describing the elastic
scattering between the solitons. The single soliton can also be called
a traveling wave solution, as its form is unchanged after some time,
up to translation \cite{AblowitzClarkson:Solitons}. Such invariances
are generalized and formalized in the symmetry approach \cite{BC74,
  Olver:LieDiff}, where one uses symmetries of the original equation
to derive an additional equation, the similarity constraint, which is
compatible with the original equation. One can then use this
constraint equation to reduce the original integrable PDE to an
integrable ordinary differential equation (ODE). For example, in the
case of the Korteweg-de Vries equation 
\begin{align}\label{kdv}
\partial_t u + \partial_x^3u+u\partial_x u=0,
\end{align}
the constraint $v\partial_x u+\partial_t u=0$ leads
to the traveling wave ansatz $u=f(x-vt)$ and elliptic equation for
$f$, while the similarity constraint $2u+x\partial_xu+3t\partial_t
u=0$ leads to the similarity ansatz $u=t^{-2/3}\phi(z),\,
z=x/(t^{1/3})$ and then to an equation for $\phi$ that can be
transformed, by letting $\phi = \partial_z y-y^2/6$, to 
\begin{equation}\label{P2}
\partial_z^2 y = \dfrac{y^3}{18} + \dfrac{yz}{3} + \alpha,
\end{equation}
(where $\alpha$ is an integration constant) which is the second Painlev\'e equation (see \cite{Olver:LieDiff} p. 195).
Given a constraint, a method for obtaining a Lax pair for the reduced equation 
was given in \cite{FlaschkaNewell}. For further applications of symmetries of PDE's
in mathematical physics see \cite{PWZ:I, PWZ:II} and references therein.

Integrable partial difference equations (P$\Delta$Es), or lattice
equations, can be seen as discrete analogues of integrable PDEs, and they have
been shown to possess many of the same characteristics as their
continuous analogues, such as Lax representations \cite{QNCvdL:NSG,
  Nijhoff:dSKdV, Nijhoff:Linearisation}, bilinear structures and
N-soliton solutions \cite{Hirota:DKdV, Hirota:DtToda,
  Hirota:dSG,Jimbo:discretesolitonI, Jimbo:discretesolitonII,
  Jimbo:discretesolitonIII}.  Furthermore, several approaches have been
developed to reduce P$\Delta$Es to ordinary difference equations
(O$\Delta$Es) \cite{dKdVreds, Quispel:Intreds, TKQ09, K09, KRQ07, KQ10, SimilarityReds, HKQT13}.
Reductions of the kind of those presented here are obtained in \cite{LeviW, QCS} using Lie group techniques in the case of
differential-difference equations.

We consider equations defined on the Cartesian two-dimensional lattice.
In this context a particularly interesting set of equations is
given by the form
\begin{equation}\label{Q}
Q(w_{l,m},w_{l+1,m},w_{l,m+1},w_{l+1,m+1};\alpha_{l},\beta_{m}) =
0,\,\forall l,m,
\end{equation}
where the subscripts, $l,m$, indicate a point in the Cartesian
2-dimensional lattice on which the dependent variable $w$ is defined,
and $\alpha_l$ and $\beta_m$ are lattice parameters associated with
the horizontal and vertical edges respectively. \footnote{We note that reductions
from P$\Delta$Es to O$\Delta$Es have also been derived for equations depending on more
  points of the lattice, as well as for systems of equations
  \cite{K09, KQ10}.}
Such equations are
often called quad equations, because the equation connects values of
$w$ given at the corners of an elementary quadrilateral of the
lattice, and if the parameters $\alpha_l$ and $\beta_m$ do not depend
on the coordinates $l,m$, respectively, then the equation is said to
be autonomous. We assume also that equation \eqref{Q} is
multilinear so that we can solve for any particular corner value in
terms of the other three.

For quadrilateral equations one definition of integrability
is by ``multidimensional consistency''  \cite{Nijhoff:dSKdVP6,NijWal}. This has
turned out to be a very effective definition, and in its
three-dimensional version (Consistency-Around-a-Cube, CAC) it has led
(under some mild additional assumptions) to a classification
of scalar integrable quadrilateral equations \cite{ABS:ListI, ABS:ListII}.
Our examples have been chosen from this class of equations.
One very important consequence of the CAC property is that it immediately
provides a Lax pair \cite{Nij02}, which is a system of linear difference equations
whose consistency is equivalent to the equation (\ref{Q}).

One may consider the analogue of a traveling wave solution to be a
solution on the lattice admitting the constraint\footnote{ Another type of reduction, via a nonlinear
  similarity constraint, was given in \cite{SimilarityReds}.}
\begin{equation}\label{eq:cons1}
w_{l+s_1, m+s_2} =w_{l,m},
\end{equation}
leading to what is known as an $(s_1,s_2)$-reduction \cite{K09}.
In order to construct consistent evolution we have to consider
initial values satisfying this constraint and make sure that
the evolution does not break the constraint.

In a similar manner to the continuous case, where reductions of PDE's lead 
to interesting ODE's, many authors have identified reductions given by \eqref{eq:cons1} 
with interesting O$\Delta$E's such as discrete analogues of elliptic 
functions, known as QRT maps \cite{SimilarityReds}, 
discrete Painlev\'e equations \cite{Gramani:Reductions, Hay, OvdKQ:reductions, Ormerod:qP6, Gramani:Q4Ell}
and many higher dimensional mappings \cite{dKdVreds, Quispel:Intreds, TKQ09, KQ10, HKQT13}. Of particular interest 
to this study are QRT maps and discrete Painlev\'e equations, which are both classes of integrable second order nonlinear difference equations. 
The QRT maps are autonomous mappings that preserve a biquadratic invariant \cite{QRT1, QRT2} whereas 
discrete Painlev\'e equations are integrable non-autonomous difference equations admitting
the classical Painlev\'e equations as continuum limits \cite{DPS} and also QRT maps
as autonomous limits \cite{ramani2002autonomous}. For example, two discretizations of \eqref{P2} are
\begin{align*}
y_{n+1} + y_{n-1} = \dfrac{y_n(hn + a)+b}{1-y_n^2},\\
y_{n+1}y_ny_{n-1} = \dfrac{aq^ny_n(y_n - q^n)}{y_n-1},
\end{align*}
which are called multiplicative and additive difference equations in accordance with their 
dependence on $n$ \cite{DPS}. Their autonomous limits, when $h \to 0$ and $q\to 1$ respectively, 
are QRT maps \cite{QRT1, QRT2}. 

Let us consider the simplest (nontrivial) case of a periodic reduction, determined by the constraint
$w_{l+1,m-1}=w_{l,m}$. We can then give the initial values on the
blue staircase given in Figure \ref{fig:intro}.\begin{figure}[ht]
\begin{tikzpicture}[scale=1.2]
\draw[black!30] (-1.5,-.5) grid (4.5,3.5);
\draw[blue,thick] (-1.5,3) -- (-1,3) -- (-1,2) -- (0,2) --
 (0,1)-- (1,1)-- (1,0)--(2,0)--(2,-0.5);
\draw[red,thick] (-1,3.5) -- (-1,3) -- (0,3) -- (0,2) -- (1,2) --
 (1,1)-- (2,1)-- (2,0)--(3,0)--(3,-0.5);
\begin{scope}[xshift=.3cm,yshift=-.3cm]
\node at (-1.45,3.15) {$y$};
\node at (-1.45,2.15) {$x$};
\node at (-0.45,2.15) {$y$};
\node at (-0.45,1.15) {$x$};
\node at (0.55,1.15) {$y$};
\node at (0.55,0.15) {$x$};
\node at (1.55,0.15) {$y$};

\node at (-0.15,3.45) {$y'$};
\node at (-0.15,2.45) {$x'$};
\node at (0.85,2.45) {$y'$};
\node at (0.85,1.45) {$x'$};
\node at (1.85,1.45) {$y'$};
\node at (1.85,0.45) {$x'$};
\node at (2.85,0.45) {$y'$};
\end{scope}
\filldraw[red] (0,3) circle (.05);
\filldraw[red] (1,2) circle (.05);
\filldraw[red] (2,1) circle (.05);
\filldraw[red] (3,0) circle (.05);
\end{tikzpicture}
\caption{Labeling of variables for the (1,-1)-reduction of the
  lattice. \label{fig:intro}}
\end{figure}
 In this case only two initial values are needed, $x$ and $y$. Solving for
 $w_{l+1,m+1}$ from \eqref{Q} we obtain
\[
w_{l+1,m+1}=f(w_{l,m},w_{l+1,m},w_{l,m+1};\alpha,\beta)
\]
for some rational function $f$ (here we assume the parameters $\alpha,\beta$
are constants). From Figure \ref{fig:intro} we then find that the
initial values on the staircase evolve by the two dimensional map
\[
x'=y,\quad y'=f(x,y,y;\alpha,\beta),
\]
and that in particular the periodicity is preserved.
This result can also be written as a second order ordinary difference equation
of the form
\[
x_{n+2}=f(x_n,x_{n+1},x_{n+1};\alpha,\beta).
\]
What is important is that if the original P$\Delta$E \eqref{Q} is
integrable and has a Lax pair then it is possible to construct a Lax
pair for the resulting ordinary difference equation, which therefore is
integrable as well. 

Recently three of the authors presented a direct method for obtaining the 
Lax representations of equations arising as periodic reductions 
of non-autonomous lattice equations \cite{OvdKQ:reductions, Ormerod:qP6}, 
which can be consider the discretization of the method given in 
\cite{FlaschkaNewell}.

In this paper we consider the generalization of \eqref{eq:cons1} in
the form
\begin{align}\label{twist}
w_{l+s_1,m+s_2} = T(w_{l,m}),
\end{align}
where the transformation $T$ (which we call the ``twist'') is
fractional linear, which is also known as a homographic transformation \cite{DuVal}. 
In the example discussed above we would impose $w_{l+1,m-1} = T(w_{l,m})$ 
and start with a sequence of initial values of the form
\[
\dots,T^{-2}(y), T^{-1}(x), T^{-1}(y), x,y,T(x),T(y),\dots
\]
and after one step of evolution the new values should be similarly
related, i.e.,
\[
\dots,T^{-2}(y'), T^{-1}(x'), T^{-1}(y'), x',y',T(x'),T(y'),\dots
\]
Thus on the $k$-th step of the staircase we would get the evolution
\[
T^{k}(x')=T^k(y),\quad T^k(y')=f(T^{k+1}(x),T^{k+1}(y),T^{k}(y)),
\]
But since $y'=f(T(x),T(y),y)$ this makes sense only if
\[
 T^k(f(T(x),T(y),y))=f(T^{k+1}(x),T^{k+1}(y),T^{k}(y)),
\]
in other words, equation  \eqref{Q} must be invariant under the
transformation $T$, i.e.
\[
Q(\{T(w_{l,m})\};\alpha_{l},\beta_{m}) \propto Q(\{w_{l,m}\};\alpha_{l},\beta_{m}) .
\]

The main result of the paper is a method for calculating Lax
representations for these reductions, even in the non-autonomous case.

The paper is organized as follows: First in \S \ref{consistency}
we review the reduction method for the $s_1=2,s_2=1$ reduction and
then discuss the possible non-autonomous parameters of the equation. We
distinguish the following cases, based on how the lattice parameters
$\alpha_l$ and $\beta_m$ vary:
\begin{itemize}
\item the autonomous case, where the parameters are constant;
\item the simply non-autonomous case, where the parameters depend only explicitly
on the lattice position; and
\item the fully non-autonomous case, where the parameters also depend
  on additional constants, which are not left invariant under a
  lattice shift.
\end{itemize}
Each of these three cases exists in a twisted and a non-twisted version.
We will review these parameter choices in more depth in \S \ref{consistency}.

In \S \ref{sec:Twist} we present the general method for
constructing the Lax matrices. To illustrate our method we then 
perform $(2,1)$-reductions of three archetypical equations with distinct
twists. The first equation of the form (\ref{Q}), considered in \S \ref{sec:dmKdV}, will be
the discrete modified Korteweg-de Vries equation (dmKdV), also called
$H3_{\delta=0}$, where
\begin{align}\label{dmKdV}
Q_{H3_{\delta=0}}=
\alpha_l(w_{l,m}&w_{l+1,m} - w_{l,m+1}w_{l+1,m+1})\\
&-\beta_m(w_{l,m}w_{l,m+1} - w_{l+1,m}w_{l+1,m+1}),\nonumber
\end{align}
with twist $T_1:w\to w\lambda$. 
Here we will review the non-twisted autonomous case, the twisted
autonomous case, the twisted simply non-autonomous case and the
twisted fully non-autonomous case. We will also briefly study a second
twist, $T_2:w\to \frac{\lambda}{w}$. 

The second equation, considered in \S \ref{sec:dpKdV}, will be the
lattice potential Korteweg-de Vries equation, or $H1$, where
\begin{equation}\label{dpKdV}
Q_{H1}=(w_{l,m}-w_{l+1,m+1})(w_{l+1,m}-w_{l,m+1}) -  \alpha_l+ \beta_m,
\end{equation}
with twists $T_1: w \to w+ \lambda$ and $T_2 : w \to \lambda - w$. 
Here we will consider the twisted autonomous case, the twisted simply
non-autonomous case and the twisted fully non-autonomous case. 
In \S \ref{sec:dSKdV}, we will consider the lattice Schwarzian
Korteweg-de Vries equation, or $Q1_{\delta=0}$, with
\begin{align}
\label{dSKdV}
Q_{Q1_{\delta=0}}= \alpha_{l}[(w_{l,m} & - w_{l,m+1})(w_{l+1,m}-w_{l+1,m+1})]\\
&-\beta_{m}[(w_{l,m} - w_{l+1,m})(w_{l,m+1}-w_{l+1,m+1})],\nonumber
\end{align}
where we will consider the twisted autonomous case and the twisted
fully non-autonomous case. The twist will be an arbitrary M\"obius
transformation.  

Finally in \S \ref{22qpvi} we will consider the (2,2)-reduction of
\eqref{dSKdV}, and obtain the full parameter $q\text{-P}_{VI}$. In \S
\ref{s1s2} we treat the general ($s_1,s_2$)-reduction, and provide a
list of twists for ABS-equations \cite{ABS:ListI,ABS:ListII}.

While this paper was being edited, the preprint \cite{HHS13} appeared
on the arXiv, which presents a twisted version of the approach in
\cite{Gramani:Q4Ell}.

\section{Symmetry invariance}\label{consistency}
\begin{figure}[!ht]
\begin{tikzpicture}[scale=1.5]
\draw[black!30] (-1.5,-.5) grid (4.5,3.5);
\draw[blue,thick] (-1.5,0) -- (0,0) -- (0,1) -- (2,1)-- (2,2)-- (4,2)-- (4,3)--(4.5,3);
\begin{scope}[xshift=.3cm,yshift=-.3cm]
\node at (0,0) {$T^pu_n$};
\node at (1,1) {$T^pu_{n+1}$};
\node at (-.1,1) {$T^{p-1}u_{n+2}$};
\node at (2,1) {$T^{p+1} u_n$};
\node at (2,2) {$T^p u_{n+2}$};
\node at (3.1,2) {$T^{p+1} u_{n+1}$};
\node at (.8,2) {$T^{p-1}u_{n+3}$};
\node at (4.1,2) {$T^{p+2} u_{n}$};
\node at (3,3) {$T^pu_{n+3}$};
\node at (4.1,3) {$T^{p+1}u_{n+2}$};
\end{scope}
\filldraw[red] (1,2) circle (.07);
\filldraw[red] (3,3) circle (.07);
\end{tikzpicture}
\caption{Labeling of variables for the (2,1)-reduction of the lattice with respect to \eqref{labelling}. \label{fig:label}}
\end{figure}
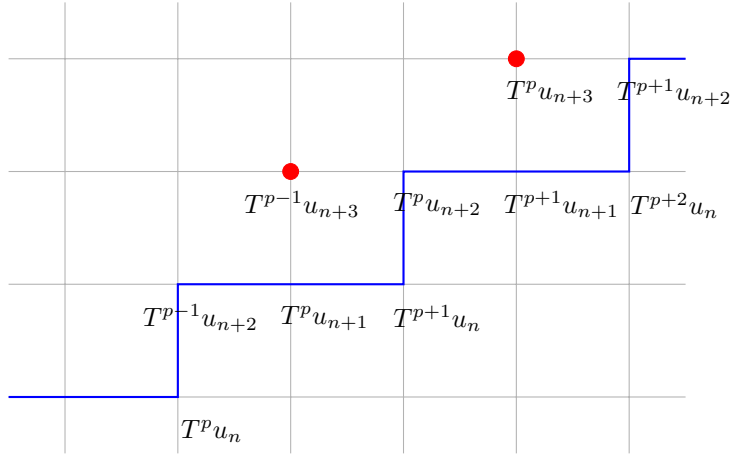

For pedagogical reasons we specialize our reduction, given by \eqref{twist}, to one of the simplest possible cases; where $s_1 =2$ and $s_2 = 1$. Contrary to the case $s_1=s_2=1$, in this case there is a difference between the simply non-autonomous case and the fully non-autonomous case.
In this special case, our reduction may be specified by introducing two variables,
\begin{equation}\label{np}
n = 2m-l, \hspace{2cm} p = l-m.
\end{equation}
We label the variables of the reduction in terms of $n$ and $p$ by specifying
\begin{equation}\label{labelling}
w_{l,m} \mapsto T^{l-m} u_{2m-l} = T^p u_n.
\end{equation}
This extends the labeling of \cite{OvdKQ:reductions} to accommodate for the twist. With this constraint, it is sufficient to specify just three initial conditions. Their values, and the values obtained from the similarity constraint, \eqref{twist}, form a staircase which determines a solution on all of $\mathbb{Z}^2$. A small portion of the staircase in $\mathbb{Z}^2$ has been depicted in Figure \ref{fig:label}.

The shift $(l,m) \to (l+1,m+1)$ leaves $p$ invariant and induces, by \eqref{np}, the shift $n \to n+1$, as one can see in Figure \ref{fig:label}.
On the top-right square in Figure \ref{fig:label} we can solve the equation,
\[
Q(T^{p+1}u_{n+1}, T^{p+2}u_n, T^{p}u_{n+3}, T^{p+1}u_{n+2};\alpha_{l+3},\beta_{m+2}) = 0,
\]
to find $u_{n+3}$, and hence the triple $(u_{n+1},u_{n+2},u_{n+3})$, from the triple $(u_{n},u_{n+1},u_{n+2})$ and the
twist $T$. But this is not the only equation for $u_{n+3}$; considering the middle square in Figure \ref{fig:label} we have
\begin{align}
\label{reductioninQ}Q(T^{p}u_{n+1}, T^{p+1}u_n, T^{p-1}u_{n+3}, T^{p}u_{n+2};\alpha_{l+1},\beta_{m+1}) = 0.
\end{align}
which may also be used to find $u_{n+3}$. In general, 
if $\alpha_{l+2}=\alpha_l$ and $\beta_{m+1}=\beta_m$, then the reduction is consistent if $T$ is chosen to be a symmetry of \eqref{Q}. In particular, if $\alpha_l = \alpha$
and $\beta_m = \beta$, are constants the resulting reductions are autonomous 3-dimensional mappings.

\vspace{.1cm}

To pass to the non-autonomous case, we notice\footnote{See Table 1 in \cite{OvdKQ:reductions} for other equations of the ABS-list admitting such a representation.} that equations \eqref{dmKdV} and \eqref{dSKdV} only depend on the
ratio $\alpha_l/\beta_m$.
For such {\em multiplicative} equations the reductions are consistent if $\alpha_{l+2}/\beta_{m+1}=\alpha_l/\beta_m$. Using separation of variables this yields
\begin{equation}\label{1storder:multfull}
\dfrac{\alpha_{l+2}}{\alpha_l} = \dfrac{\beta_{m+1}}{\beta_m}  := q^2,
\end{equation}
which is a second order equation in  $\alpha_l$ and first order in $\beta_m$. The general fully non-autonomous solution to \eqref{1storder:multfull} is 
\begin{equation}\label{H3Q1fullvars}
\alpha_l = \left\{ \begin{array}{c p{2cm}} 
a_0 q^l & if $l$ is even,\\
a_1 q^l & if $l$ is odd,\end{array}\right. \hspace{1cm} \beta_m = b_0q^{2m},
\end{equation}
where we may absorb $b_0$ in $a_0,a_1$, or simply take $b_0=1$. The resulting reduction may be expressed in terms of
$\beta_m/\alpha_l  \propto q^n$. 

Equation \eqref{dpKdV} may be written explicitly as a function of $\alpha_l-\beta_m$. For such {\em additive} equations, separation of variables yields
\begin{equation}\label{1storder:addfull}
\alpha_{l+2} - \alpha_l = \beta_{m+1} - \beta_m := 2h.
\end{equation}
The general fully non-autonomous solution to \eqref{1storder:addfull} is
\begin{equation}\label{H1fullvars}
\alpha_l = \left\{ \begin{array}{c p{2cm}} 
a_0 + lh & if $l$ is even,\\
a_1 + lh & if $l$ is odd,\end{array}\right. \hspace{1cm} \beta_m = b_0 + 2hm.
\end{equation}
Here we may, without loss of generality, take $b_0=0$. In the additive case, the reduction will depend on $\alpha_l-\beta_m$ which depends linearly on the variable $n=2m-l$.

For both these additive and multiplicative equations, the special reductions where $a_i$ and $b_i$ do not
depend on $i$ will be called simply non-autonomous. For the fully non-autonomous reductions the shift $n \to n+1$
has the effect of swapping the roles of $a_0$ and $a_1$. We have two options here: either to introduce
a second root of unity, or to consider the second iterate of the map. We choose the second option in this paper.

\section{Twist Matrices and Lax representations} \label{sec:Twist}
In this section we will provide a method to construct Lax representations for twisted reductions. Firstly, let us consider a Lax pair for a lattice equation given by a pair of linear difference equations 
\begin{subequations}\label{LaxQ}
\begin{align}
\label{LaxL}\Psi_{l+1,m}(\gamma) &= L_{l,m}(\gamma) \Psi_{l,m}(\gamma),\\
\label{LaxM}\Psi_{l,m+1}(\gamma) &= M_{l,m}(\gamma) \Psi_{l,m}(\gamma),
\end{align}
\end{subequations}
where $\gamma$ is a spectral parameter.
This is a Lax pair in the sense that the compatibility condition between \eqref{LaxL} and \eqref{LaxM}, which can be written as
\begin{equation} \label{CompLM}
L_{l,m+1}M_{l,m} - M_{l+1,m}L_{l,m} = 0,
\end{equation} 
is equivalent to imposing \eqref{Q}. For 3D-consistent equations of the form \eqref{Q}, cf. \cite{ABS:ListI, ABS:ListII}, the matrices $L_{l,m}$ and
$M_{l,m}$ are explicitly given in terms of derivatives of the function $Q$ \cite[Equation 1.10]{OvdKQ:reductions}. Therefore, and the importance of this will be apparent later on, because \eqref{dmKdV} and \eqref{dSKdV} are functions of $\alpha_l/\beta_m$, the Lax matrices for \eqref{dmKdV} and \eqref{dSKdV}, $L_{l,m}$ and $M_{l,m}$, will be functions of $\alpha_l/\gamma$ and $\beta_m/\gamma$ respectively. Similarly the Lax matrices $L_{l,m}$ and $M_{l,m}$ for \eqref{dpKdV} are functions of $\alpha_l - \gamma$ and $\beta_m - \gamma$ respectively. 

To arrive at a particular form of the Lax pairs, we will sometimes perform a gauge transformation. 
For example, if the reduced equations can be dimensionally reduced by choosing
special variables, one would like to also express the Lax pair in terms of these variables.
Then one considers 
\begin{equation}\label{gauge}
\Psi_{l,m}' = Z_{l,m} \Psi_{l,m}.
\end{equation} 
The linear system satisfied by $\Psi_{l,m}'$ is 
\begin{subequations}\label{guagetransformedLax}
\begin{align}
\Psi_{l+1,m}' = (Z_{l+1,m} L_{l,m} Z_{l,m}^{-1})\Psi_{l,m}' = L_{l,m}'\Psi_{l,m}',\\
\Psi_{l,m+1}' = (Z_{l,m+1} M_{l,m} Z_{l,m}^{-1}) \Psi_{l,m}'= M_{l,m}'\Psi_{l,m}'.
\end{align}
\end{subequations}
In a slight abuse of notation, we will not distinguish between the pair $(L_{l,m},M_{l,m})$ and $(L_{l,m}',M_{l,m}')$.

Before we turn to the key ansatz we make in order to derive Lax pairs
for the reduction, one must realise that if a solution to the lattice equation is known,
$w_{l,m}$ for all $l,m \in \mathbb{Z}$, one can obtain a fundamental solution of the linear problem.
Relating the behaviour of solutions of the nonlinear partial differential equation with its spectral problem 
plays a fundamental role in inverse scattering methods for partial differential equations \cite{AblowitzSegur}.
The discrete analogue of this theory for systems of difference equations has also been studied \cite{DISPI, DISPII} 
and applied to a system of the form \eqref{Q} by Butler et al. \cite{Butler:scatteringKdV}. Our key anstatz is based 
on a relation between the solutions of systems defined by \eqref{Q} and solutions of \eqref{LaxQ}.

Let us start with the autonomous case. Given the fact that any solution, $w_{l,m}$, lifts to a solution of the linear problem, we may lift a solution satisfying \eqref{twist} to a system that is now dependent on the variables $u_n$. That is to say, we have a solution to some linear system
\[
\Psi_{l,m}(\gamma;\{w_{l,m}\}) \mapsto Y_n(\gamma;\{u_n\}).
\]
We proceed as per usual, and construct operators, $A_n$ and $B_n$, which are equivalent to shifts in $l$ and $m$ given by $(l,m) \to (l+2,m+1)$, and $(l,m) \to (l+1,m+1)$, respectively. These are given by the products
\begin{subequations}
\begin{align}
\label{prodA}A_n(\gamma) &\mapsfrom L_{l+1,m+1}L_{l,m+1}M_{l,m},\\
\label{prodB}B_n(\gamma) &\mapsfrom L_{l,m+1}M_{l,m}.
\end{align}
\end{subequations}
The matrix $A_n$ is called the monodromy matrix. It corresponds to a path, in Figure \ref{fig:label}, from $u_n$ to $Tu_n$, going up one step and to the right two steps. We note that, in general, the matrix $B_n$ is a particular factor of $A_n$, namely the one that corresponds to
the shift $n\mapsto n+1$. The function $Y_n(\gamma;\{u_n\})$ now satisfies the equation
\begin{subequations}
\begin{align}
\label{defA}T Y_{n}(\gamma;\{u_n\}) &= A_n(\gamma) Y_n(\gamma;\{u_n\})\\
\label{defB}Y_{n+1}(\gamma;\{u_n\}) &= B_n(\gamma) Y_n(\gamma;\{u_n\}),
\end{align}
\end{subequations}
where, from the above, we may lift our symmetry, $T$, to the level of the linear problem via application on the $w_{l,m}$ (or equivalently on $u_n$).

Our {\em key ansatz} is that there is the additional relation 
\begin{equation}\label{twistmatrix} 
Y_n(\gamma;\{T u_n\}) Y_n(\gamma;\{u_n\})^{-1} = S_n(\{u_n\}),
\end{equation}
where $S_n$ does not depend on the spectral parameter. This rather innocuous looking relation implies that the singularities of $Y_n$, as a function of the spectral parameter $\gamma$, are independent of any particular solution of the lattice equation. That is, the singularities and poles of $Y_n(\gamma;\{T u_n\})$ are cancelled out by the poles and singularities of $Y_n(\gamma;\{u_n\})^{-1}$ to give a constant matrix, $S_n(\{u_n\})$, which we call the {\em twist matrix}. For all examples of twisted reductions provided, we have been able to obtain such twist matrices.

Now, combining the two equations \eqref{defA} and \eqref{twistmatrix} we obtain the first half of a standard Lax pair for an autonomous mapping
\begin{align}
\label{DSA}Y_n(\gamma) = S_n^{-1} A_n(\gamma) Y_n(\gamma).
\end{align}
where the other half of the Lax pair is \eqref{defB}. The compatibility between \eqref{DSA} and \eqref{defB}, which is equivalent to the autonomous reduction, is given by
\begin{equation}\label{autcomp}
S_{n+1}^{-1} A_{n+1}(\gamma) B_n(\gamma) - B_n(\gamma) S_n^{-1} A_n(\gamma) = 0,
\end{equation}
and integrals for this reduction can be obtained by taking the trace of the twisted monodromy matrix $S_n^{-1} A_n(\gamma)$.

While $A_n$ and $B_n$ are determined by \eqref{prodA} and \eqref{prodB}, the task of determining $S_n$ remains. As is typical in integrable systems, the linear system is overdetermined, which gives us a straightforward, albeit complicated, way of calculating $S_n$. The complication arises because one needs to simultaneously calculate the twist matrix and the evolution equation from the compatibility condition, thereby increasing the number of conditions that need to be satisfied without increasing the number of relations from the compatibility. However, there is a simpler way to calculate $S_n$; observe that when we use \eqref{twistmatrix}, \eqref{defA} and \eqref{defB}, we get
\begin{align*}
T Y_{n+1} = T(B_n)A_n Y_n = A_{n+1} B_n Y_n.
\end{align*}
Rewriting \eqref{autcomp} yields the relation
\[
A_{n+1} B_n = S_{n+1} B_n S_n^{-1} A_n.
\]
By combining these equations, and by cancelling irrelevant factors, we obtain
\begin{equation}\label{calcS}
T(B_n)S_n = S_{n+1} B_n,
\end{equation}
which gives us an elegant way of calculating the twist matrix $S_n$ and $S_{n+1}$, that does not rely explicitly on using the reduction. We will see that
for the examples provided, the twist matrices are actually quite succinct. Furthermore, they tend to the identity matrix in the limit where the twists tends to the identity transformation.\footnote{Note that we also consider some examples of twists that are not homotopic to the identity transformation.}

The non-autonomous case is a simple generalisation of the above, since nothing we did relied upon any of the properties of $\alpha_l$ or $\beta_m$. We just need to specify a new spectral parameter for our reduced system. For the multiplicative equations, \eqref{dmKdV} and \eqref{dSKdV}, we know that the $L_{l,m}$ and $M_{l,m}$ matrices are functions of $\alpha_l/\gamma$ and $\beta_m/\gamma$ respectively, which for our choices of parameters \eqref{H3Q1fullvars}, can both be written in terms of $q^l/\gamma$ and $q^n$ only. This motivates the choice
\begin{equation}\label{spectralchoice}
x = q^l/\gamma,
\end{equation}
as our spectral parameter. This implies that the shifts $(l,m) \to (l+2,m+1)$ and $(l,m) \to (l+1,m+1)$ both have the effect of translating $x$. As in the autonomous case, we may write $A_n(x)$ and $B_n(x)$ as products of matrices $L_{l,m}$ and $M_{l,m}$:
\begin{subequations}\label{nonautABprod}
\begin{align}
\label{prodnA}A_n(x) &\mapsfrom L_{l+1,m+1}L_{l,m+1}M_{l,m},\\
\label{prodnB}B_n(x) &\mapsfrom L_{l,m+1}M_{l,m}.
\end{align}
\end{subequations}
where the linear problem, which is now in $x$, satisfies the equations
\begin{subequations}\label{LaxYABm}
\begin{align}
\label{defnA}T Y_n(q^2x) = A_n(x) Y_n(x),\\
\label{defnB}Y_{n+1}(qx) = B_n(x) Y_n(x),
\end{align}
\end{subequations}
where, for the same reasons as above, we have the additional relation
\begin{equation}\label{nonauttwistmatrix} 
T Y_n(x) = S_n Y_n(x).
\end{equation}
This means our compatibility may be written
\begin{equation}\label{compnonautmult}
S_{n+1}^{-1} A_{n+1}(qx) B_n(x) - B_n(q^2x) S_n^{-1} A_n(x) = 0,
\end{equation}
where $S_n$ is actually the same twist matrix as in the autonomous case. 

For the additive equation, \eqref{dpKdV}, the Lax matrices, $L_{l,m}$ and $M_{l,m}$, are functions of $\alpha_l - \gamma$ and $\beta_m-\gamma$ respectively. For the non-autonomous parameter choice, \eqref{H1fullvars}, $L_{l,m}$ and $M_{l,m}$ are both functions of $hl-\gamma$ and $nh$. This motivates the definition 
\begin{equation}\label{spectralchoiceadd}
x = hl - \gamma.
\end{equation}
Using the same product formulas for $A_n(x)$ and $B_n(x)$, given by \eqref{nonautABprod}, the matrix $Y_n(x)$ satisfies the equations
\begin{subequations}
\begin{align}
\label{defaA}T Y_n(x+2h) = A_n(x) Y_n(x),\\
\label{defaB}Y_{n+1}(x+h) = B_n(x) Y_n(x),
\end{align}
\end{subequations}
and also equation \eqref{nonauttwistmatrix}. This means that the compatibility yields
\begin{equation}\label{compnonautadd}
S_{n+1}^{-1} A_{n+1}(x+h) B_n(x) - B_n(x+2h) S_n^{-1} A_n(x) = 0.
\end{equation}
In the following three sections we will provide the details for the (2,1)-reductions of our three examples to demonstrate this theory. We postpone the theory for general ($s_1,s_2$)-reduction to \S \ref{s1s2}.

\section{Reductions of the lattice modified Korteweg-de Vries equation}\label{sec:dmKdV}

The discrete modified Korteweg-de Vries equation (aka $H3_{\delta =0}$) , given by \eqref{dmKdV}, was one of the earliest known integrable lattice equations. It appeared as a discrete analogue of the sine-Gordon equation (equivalent under a transformation) in the work of Hirota \cite{Hirota:dSG} and its Lax pair was derived using direct linearisation \cite{QNCvdL:NSG}. Reductions of this equation have been considered by many authors \cite{SimilarityReds, Quispel:Intreds, Gramani:Reductions, TKQ09, KRQ07, HKQT13, Hay, Ormerod:qP6}. The equation has a Lax representation given by \eqref{LaxQ} where the Lax matrices are
\begin{subequations}\label{LaxH3}
\begin{align}
L_{l,m} (\alpha_l/\gamma)=& \begin{pmatrix}
 \dfrac{\gamma }{\alpha_l} & w_{l+1,m} \\[10pt]
 \dfrac{1}{w_{l,m}} & \dfrac{\gamma  w_{l+1,m}}{w_{l,m} \alpha_l}
 \end{pmatrix} ,\\
 M_{l,m} (\beta_m/\gamma)=& \begin{pmatrix}
 \dfrac{\gamma }{\beta_m} & w_{l,m+1} \\[10pt]
 \dfrac{1}{w_{l,m}} & \dfrac{\gamma  w_{l,m+1}}{w_{l,m} \beta_m}
 \end{pmatrix}.
\end{align}
\end{subequations}

We will first recall how the autonomous periodic reductions are obtained, then proceed to generalise the reductions and their Lax representations to the twisted, and non-autonomous cases. In the periodic case, \eqref{twist} is still valid, as is all the theory contained in \S \ref{consistency} and \S \ref{sec:Twist}, with the specialisation to $T(w_{l,m}) = w_{l,m}$. This means that the labelling, \eqref{labelling}, is simply
\[
w_{l,m} \mapsto u_{2m-l} = u_n.
\]
In this case, the equation governing the reduction, \eqref{reductioninQ}, is given by
\begin{equation}\label{H1autountwisted}
u_{n+3} = \dfrac{u_n \left(\alpha  u_{n+1}+\beta  u_{n+2}\right)}{\alpha  u_{n+2}+\beta  u_{n+1}}
\end{equation}
Using \eqref{prodA} and \eqref{prodB} we obtain the two Lax matrices, $A_n$ and $B_n$, given by
\begin{align*}
A_n(\gamma) =& \begin{pmatrix} \dfrac{\gamma}{\alpha} & u_n \\[10pt]
\dfrac{1}{u_{n+1}} & \dfrac{\gamma u_n}{\alpha u_{n+1}}\end{pmatrix}
B_n(\gamma),\\
B_n(\gamma) =& \begin{pmatrix} \dfrac{\gamma}{\alpha} & u_{n+1} \\[10pt]
\dfrac{1}{u_{n+2}} & \dfrac{\gamma u_{n+1}}{\alpha u_{n+2}}\end{pmatrix}
\begin{pmatrix} \dfrac{\gamma}{\beta} & u_{n+2} \\[10pt]
\dfrac{1}{u_{n}} & \dfrac{\gamma  u_{n+2}}{\beta u_{n}}\end{pmatrix}.
\end{align*}
We note that since $T(B_n) = B_n$, equation \ref{calcS} becomes $B_nS_n=B_nS_{n+1}$, where $S_n$ is a priori unknown. We parameterize $S_n$ by letting
\begin{equation}\label{arbitrarySn}
S_n = \begin{pmatrix} s_{1,n} & s_{2,n} \\ s_{3,n} & s_{4,n} \end{pmatrix}.
\end{equation}
At the coefficient of $\gamma^2$, we obtain
\begin{align*}
s_{1,n+1} = s_{1,n}, \hspace{.5cm} u_n s_{2,n} = u_{n+1}s_{2,n+1}, \hspace{.5cm} u_{n+1} s_{3,n} = u_n s_{3,n+1} , \hspace{.5cm} s_{4,n+1} = s_{4,n},
\end{align*}
and at the coefficient of $\gamma$, we then obtain
\[
s_{1,n} = s_{4,n}, \hspace{1cm} s_{2,n} = 0 , \hspace{1cm} s_{3,n} = 0.
\]
This tells us that we may choose $S_n = I$. Thus the twisted monodromy matrix coincides with the standard monodromy matrix, which shouldn't come as a surprise. Taking the trace of the
monodromy matrix gives us $\alpha\beta\Tr(A_n)=\frac{2}{\alpha}\gamma^3+K_{\eqref{H1autountwisted}}\gamma$ where
\[
K_{\eqref{H1autountwisted}}
=\alpha\left(\frac{u_n}{u_{n+2}}+\frac{u_{n+2}}{u_n}\right)
+\beta\left(\frac{u_n}{u_{n+1}}+\frac{u_{n+1}}{u_n}+\frac{u_{n+1}}{u_{n+2}}+\frac{u_{n+2}}{u_{n+1}}\right)
\]
is an integral, or constant of motion, of equation \eqref{H1autountwisted}.
One can verify that \eqref{autcomp} is satisfied on solutions of  \eqref{H1autountwisted}. 
This equation, under the transformation $y_n = u_{n+1}/u_n$, takes the more familiar form of a second order difference equation
\begin{equation} \label{mfso}
y_{n+1} y_n y_{n-1} = \dfrac{\alpha + \beta y_n}{\beta + \alpha y_n},
\end{equation}
which is more clearly a mapping of QRT type \cite{QRT1, QRT2}. The integral $K_{\eqref{H1autountwisted}}$ is also invariant under scaling and hence can be also written in terms of the reduced variable $y_n$,
\[
K_{\eqref{mfso}}
=\alpha\left(y_ny_{n+1}+\frac{1}{y_ny_{n+1}}\right) +\beta\left(y_n+\frac{1}{y_n}+y_{n+1}+\frac{1}{y_{n+1}}\right).
\]
This reduction appeared in \cite{Quispel:Intreds}. We will now give a one-parameter integrable generalisation of this reduction by considering the twisted case. 

\vspace{.1cm}

The twist we apply is given by $T(w_{l,m}) = \lambda w_{l,m}$, which means that
\[
w_{l,m} \mapsto \lambda^{l-m} u_{2m-l} = \lambda^p u_n.
\]
Under this identification, the reduction, \eqref{reductioninQ}, is given by
\begin{align}\label{QRT2}
u_{n+3} = \dfrac{\lambda ^2 u_n \left(\alpha  u_{n+1}+\beta  u_{n+2}\right)}{\alpha  u_{n+2}+\beta  u_{n+1}},
\end{align}
To obtain a Lax pair, we construct the operators $A_n$ and $B_n$, using the product representation, \eqref{prodA} and \eqref{prodB}, to give
\begin{align*}
A_n(\gamma) =& \begin{pmatrix} \dfrac{\gamma}{\alpha} & \lambda u_n \\[10pt]
\dfrac{1}{u_{n+1}} & \dfrac{\gamma \lambda u_n}{\alpha u_{n+1}}\end{pmatrix}
B_n(\gamma),\\
B_n(\gamma) =& \begin{pmatrix} \dfrac{\gamma}{\alpha} & u_{n+1} \\[10pt]
\dfrac{\lambda}{u_{n+2}} & \dfrac{\gamma \lambda u_{n+1}}{\alpha u_{n+2}}\end{pmatrix}
\begin{pmatrix} \dfrac{\gamma}{\beta} & \dfrac{u_{n+2}}{\lambda} \\[10pt]
\dfrac{1}{u_{n}} & \dfrac{\gamma  u_{n+2}}{\beta \lambda u_{n}}\end{pmatrix}.
\end{align*}
We play the same game, where $S_n$ is a priori unknown, hence, we let $S_n$ be given by \eqref{arbitrarySn}. The coefficient of $\gamma^2$ in \eqref{calcS} gives us the same conditions as in the periodic case, and at the coefficient of $\gamma$ we find
\[
s_{1,n} = \lambda s_{4,n}, \hspace{1cm} s_{2,n} = 0 , \hspace{1cm} s_{3,n} = 0.
\]
This gives us our first non-trivial twist matrix, given by
\begin{equation}\label{H3Sn}
S_n = \begin{pmatrix} \lambda & 0 \\ 0 & 1 \end{pmatrix}.
\end{equation}
Taking the trace of the twisted monodromy matrix gives us
\[
\alpha\beta\Tr(S_n^{-1}A_n)=\left(\frac{1}{\lambda\alpha}+\frac{\lambda}{\alpha}\right)\gamma^3+K_{\eqref{QRT2}}\gamma,
\]
where
\[
K_{\eqref{QRT2}}
=\alpha\left(\frac{\lambda u_n}{u_{n+2}}+\frac{u_{n+2}}{\lambda u_n}\right)
+\beta\left(\frac{\lambda u_n}{u_{n+1}}+\frac{u_{n+1}}{\lambda u_n}+\frac{\lambda u_{n+1}}{u_{n+2}}+\frac{u_{n+2}}{\lambda u_{n+1}}\right)
\]
is an integral for \eqref{QRT2}.
In the limit as $\lambda \to 1$, we retrieve the periodic case, making this a nice one-parameter family of reductions and their Lax pairs and integrals. This provides all the elements for \eqref{autcomp} to give \eqref{QRT2}. Once again, by identifying $y_n = u_{n+2}/u_{n+1}$, we have the classic QRT map
\begin{equation} \label{cqrt}
y_{n+1}y_ny_{n-1} = \dfrac{\lambda^2 (\alpha + \beta y_n)}{\beta + \alpha y_n},
\end{equation}
with corresponding integral obtained from $K_{\eqref{QRT2}}$. Thus we have obtained a one parameter generalisation of the reduction, \eqref{H1autountwisted}, found in \cite{Quispel:Intreds}.

\vspace{.1cm}

When we turn to the simply non-autonomous case, we obtain a version of $q$-$\mathrm{P}_{II}$. In taking $\alpha_l=aq^l$ and $\beta_m=q^{2m}$, we need to take into account the position of the square we use to evaluate the reduction. With respect to Figure \ref{labelling}, if the square whose lower left entry is $u_n$ denotes $(l,m)$, the relevant square used for \eqref{reductioninQ} is at $(l+1,m+1)$ Thus we obtain the reduction
\begin{equation}\label{qP2u}
u_{n+3}=\dfrac{\lambda ^2 u_n \left(a u_{n+1}+ q^{n+1} u_{n+2}\right)}{a u_{n+2}+ q^{n+1} u_{n+1}}.
\end{equation}
We now use \eqref{spectralchoice} in our product representation for $A_n(x)$ and $B_n(x)$, to obtain
\begin{align*}
A_n(x) &= \begin{pmatrix} \dfrac{1}{qxa} & \lambda u_n \\[10pt]
\dfrac{1}{u_{n+1}}  & \dfrac{\lambda u_n }{qxau_{n+1}} \end{pmatrix}
B_n(x) ,\\
B_n(x) &= \begin{pmatrix} \dfrac{1}{xa} & u_{n+1} \\[10pt]
\dfrac{\lambda}{u_{n+2}}  & \dfrac{\lambda u_{n+1} }{xau_{n+2}} \end{pmatrix}
\begin{pmatrix} \dfrac{1}{xq^n} & \dfrac{u_{n+2}}{\lambda} \\[10pt]
\dfrac{1}{u_{n}}  & \dfrac{u_{n+2} }{x\lambda q^n u_{n}} \end{pmatrix}.
\end{align*}
We use the form \eqref{arbitrarySn} once more, and the calculations follow analogously to the previous case and give \eqref{H3Sn}. With $A_n(x)$, $B_n(x)$ and $S_n$ defined, the compatibility, \eqref{compnonautmult}, gives \eqref{qP2u}. Furthermore, by letting $y_n = u_{n+2}/u_{n+1}$, we find a more direct correspondence with a $q$-analogue of the second Painlev\'e equation found in \cite{Gramani:coalescences},
\begin{equation}
y_{n+1}y_ny_{n-1} = \dfrac{\lambda^2 (a + q^{n+1} y_n)}{q^{n+1} + a y_n},
\end{equation}
which generalizes a reduction of Nijhoff and Papageorgiou \cite{SimilarityReds}. At this point, we note that we may use alternative Lax matrices to \eqref{LaxH3}. By considering a transformation of the form \eqref{gauge}, where
\[
Z_{l,m} = \begin{pmatrix} \dfrac{1}{w_{l,m}} & 0 \\ 0 & 1 \end{pmatrix}
\]
we obtain a Lax pair given by 
\begin{align}
L_{l,m} =& \begin{pmatrix}
 \dfrac{\gamma w_{l,m}}{\alpha_l w_{l+1,m}} & 1 \\[10pt]
 1 & \dfrac{\gamma w_{l+1,m}}{\alpha_l w_{l,m}}
 \end{pmatrix} ,\\
 M_{l,m} =& \begin{pmatrix}
 \dfrac{\gamma w_{l,m}}{\beta_m w_{l,m+1}} & 1 \\[10pt]
 1 & \dfrac{\gamma w_{l,m+1}}{\beta_m w_{l,m}}
 \end{pmatrix} .
\end{align}
Notice that these matrices are actually invariant under the uniform application of the transformation $w_{l,m} \to T(w_{l,m})$. Since all the variables $w_{l,m}$ in $L_{l,m}$ and $M_{l,m}$ appear in ratios, the Lax pair may be expressed in the variables $y_n = u_{n+2}/u_{n+1}$. In this light, we write an alternative set of Lax matrices
\begin{align*}
A_n(x) =& \begin{pmatrix} 
\dfrac{y_{n-1}}{qx\lambda a} & 1 \\[10pt] 1 & \dfrac{\lambda}{qxa y_{n-1}} \end{pmatrix}
B_{n}(x),\\
B_{n}(x) =& \begin{pmatrix} 
\dfrac{y_{n}}{x\lambda a} & 1 \\[10pt] 1 & \dfrac{\lambda}{xa y_{n}} \end{pmatrix}
\begin{pmatrix} 
\dfrac{\lambda}{q^nxy_n y_{n-1}} & 1 \\[10pt] 1 & \dfrac{y_n y_{n-1}}{q^nx\lambda} \end{pmatrix},
\end{align*}
and twist matrix $S_n = I$. This is an immediate consequence of the fact that $y_n$ is an invariant of $T$: $T(y_n) = T(u_{n+2})/T(u_{n+1}) = u_{n+2}/u_{n+1} = y_n$.

\vspace{.1cm}

The last case to do is the fully non-autonomous generalisation, where $\alpha_l$ and $\beta_m$ are given by \eqref{H3Q1fullvars} with $b_0=1$. It should be noted that the resulting equation governing $n \to n+1$ turns an even $l$ into an odd $l$, hence, the evolution equation incorporates a change in $a_0$ and $a_1$. With this in mind, the evolution equation is given by \eqref{reductioninQ} combined with a change in $a_0$ and $a_1$: in the case that $n$ (and hence, $l$) is even, $u_{n+3}$ is calculated from 
\begin{align}\label{H3singleev}
u_{n+3} = \dfrac{\lambda ^2 u_n \left(a_1 u_{n+1}+q^{n+2} u_{n+2}\right)}{a_1 u_{n+2}+q^{n+2} u_{n+1}}, \hspace{1cm} a_0 \to \dfrac{a_1}{q}, \hspace{1cm} a_1 \to q a_0.
\end{align}
This system possesses a Lax pair of the form \eqref{LaxYABm}, where the Lax matrices are given by products \eqref{prodnA} and \eqref{prodnB}. The shift $n \to n+2$ has an alternative deformation matrix, given by $B_n(x) \mapsfrom M_{l,m}$, which simplifies the calculation. If we let $A_n(x)$ be given
by the product \eqref{prodnA}, we obtain 
\begin{align*}
A_n(x) &= \begin{pmatrix} \dfrac{1}{xa_1} & \lambda u_n \\[10pt] \dfrac{1}{u_{n+1}}  & \dfrac{\lambda u_n }{xa_1u_{n+1}} \end{pmatrix}
\begin{pmatrix} \dfrac{1}{xa_0} & u_{n+1} \\[10pt] \dfrac{\lambda}{u_{n+2}}  & \dfrac{\lambda u_{n+1} }{xa_0u_{n+2}} \end{pmatrix}
B_n(x) ,\\
B_n(x) &= \begin{pmatrix} \dfrac{1}{xq^n} & \dfrac{u_{n+2}}{\lambda} \\[10pt] \dfrac{1}{u_{n}}  & \dfrac{u_{n+2} }{x\lambda q^n u_{n}} \end{pmatrix}.
\end{align*}
By using \eqref{arbitrarySn} and \eqref{calcS} we once again obtain \eqref{H3Sn}. Here the compatibility condition is
\begin{equation}\label{compnonautmult2}
S_{n+1}^{-1} A_{n+1}(x) B_n(x) - B_n(q^2x) S_n^{-1} A_n(x) = 0,
\end{equation}
which we use to obtain \eqref{H3singleev}. However, this is not as obviously a two dimensional mapping. We employ a technique used in \cite{OvdKQ:reductions} to rewrite this equation. We take $A_n(x)$ and evaluate the root of the upper right entry (in $x^2$), denoting this $y$. The determinant of $A_n(\sqrt{y})$ factors nicely, and the factors of $\det A_n(\sqrt{y})$ appear in the diagonal entries, in addition to a simple multiplicative factor, which we denote $z_n$. Explicitly, modulo some scaling, these variables are
\begin{subequations}
\begin{align}
y_n =&   \dfrac{a_1u_{n+1}}{u_n} + \dfrac{a_0 u_{n+1}u_{n+2}+ q^n u_n u_{n+2} }{ \lambda^2  u_n^2} ,\\
z_n =&u_{n+1}\left(\dfrac{a_1\lambda u_{n+1}}{u_{n+2}}+\dfrac{a_0}{\lambda u_n}\right),
\end{align}
\end{subequations}
which then satisfy the difference equations
\begin{subequations}\label{H3nonautred}
\begin{align}
y_n y_{n+2}=&\left(\lambda  q^{n+2}+z_n\right) \left(\dfrac{q^n}{\lambda}+ 
   z_n\right)\\
   z_n z_{n+2}=&\dfrac{\left(a_1 q^{n+2}+a_0 y_{n+2}\right) \left(a_0 q^{n+2}+a_1 y_{n+2}\right)}{\left(a_0 a_1+ q^{n+2} y_{n+2}\right)}.
\end{align}
\end{subequations}
This equation first appeared in the work of Ramani et al. \cite{Quadratic} and is related, via a Miura transformation, to a version of $q$-$\mathrm{P}_{III}$ found in \cite{AsymmetricdPs}. This equation has a symmetry group which is of affine Weyl type $A_2^{(1)} + A_1^{(1)}$ \cite{Hunting, Sakai:rational}.

\vspace{.1cm}

Another possible choice of twist is $T_2: w \to \lambda/w$, which is is not homotopic to the identity.
The twist matrix associated with ($s_1,s_2)$-reductions
of \eqref{dmKdV} with fixed Lax representation, \eqref{LaxH3}, is
\[
S_n = \begin{pmatrix} 0 & \lambda \\ 1 & 0 \end{pmatrix},
\]
for a large class of $s_1$ and $s_2$. We remark that
twist matrices are not gauge invariant.

\section{Reductions of the lattice potential Korteweg-de Vries equation}\label{sec:dpKdV}

The lattice potential Korteweg-de Vries equation (aka $H1$), \eqref{dpKdV}, was derived from the direct linearisation approach \cite{Nijhoff:dSKdV}, and it yields the potential Korteweg-de Vries equation in a continuum limit. Periodic reductions of \eqref{dpKdV} were considered by many authors \cite{dKdVreds, SimilarityReds, Quispel:Intreds,Nijhoff:dSKdVP6,KQ10, HKQT13}. The $(2,1)$-periodic non-autonomous reduction and its Lax pair were recently given in \cite{OvdKQ:reductions}.

The equation \eqref{dpKdV} has a Lax representation given by \eqref{LaxQ} where the Lax matrices are given by
\begin{subequations} \label{LaxH1}
\begin{align}
L_{l,m}=& \begin{pmatrix} w_{l,m} & -\gamma + \alpha_l - w_{l,m} w_{l+1,m} \\ 
1 & - w_{l+1,m} 
\end{pmatrix},\\
M_{l,m} =& \begin{pmatrix} w_{l,m} & -\gamma + \beta_m - w_{l,m} w_{l,m+1} \\ 
1 & - w_{l,m+1} 
\end{pmatrix}.
\end{align} 
\end{subequations}

The twist that we seek to apply is the transformation $T(w_{l,m}) = w_{l,m} + \lambda$, which means our reduced variables are specified by
\[
w_{l,m} \mapsto u_{2m-l} + (l-m)\lambda =  u_n + p\lambda.
\]
For the twisted autonomous case, where $\alpha_l = \alpha$ and $\beta_m = \beta$ are constants, it is clear that we obtain the difference equation
\begin{equation}\label{QRT1}
(u_{n} - u_{n+3} + 2\lambda)(u_{n+1}-u_{n+2}) = \alpha-\beta.
\end{equation}
The Lax pair for this autonomous equation may be specified by \eqref{prodA} and \eqref{prodB}, where the lattice variables take on their reduced values, giving
\begin{align*}
A_n(\gamma) =& \begin{pmatrix} 
u_{n+1} & \alpha - \gamma - (\lambda+u_{n})u_{n+1}\\
1 & -(\lambda+u_n)
\end{pmatrix}B_n(\gamma),\\
B_n(\gamma) =& \begin{pmatrix} 
-\lambda + u_{n+2} & \alpha - \gamma - (u_{n+2}-\lambda)u_{n+1}\\
1 & -u_{n+1}
\end{pmatrix}\begin{pmatrix}
u_n & \beta - \gamma - (u_{n+2}-\lambda)u_n\\
1 & \lambda - u_{n+2}
\end{pmatrix}.
\end{align*}
We now need to calculate $S_n$, which is once again, a priori, an unknown function of $n$, hence, we label the elements of $S_n$ by \eqref{arbitrarySn}. By utilizing \eqref{calcS}, at the level of the coefficient of $\gamma$, we find $S_{n+1} = S_n$. Solving for the constant coefficient of \eqref{calcS} gives us that if $S_n$ is independent of $n$, then $s_{3,n} =0$ and $s_{2,n} = \lambda s_{1,n}$ and $s_{4,n} = s_{1,n}$, which we may simplify to give the second non-trivial twist matrix in this study, given by
\begin{equation}\label{H1Sn}
S_n = \begin{pmatrix} 
1 & \lambda \\
0 & 1 
\end{pmatrix} .
\end{equation}
Knowing $A_n(\gamma)$, $B_n(\gamma)$ and $S_n$ gives us all the necessary ingredients for calculating the compatibility, \eqref{autcomp}, which gives us the required mapping, \eqref{QRT1}. Calculating the trace of the twisted monodromy matrix,
$\Tr(S_n^{-1}A_n)=2\lambda\gamma+K_{\eqref{QRT1}}$, we obtain an integral,
\begin{align}
K_{\eqref{QRT1}}
=\alpha(u_n-u_{n+2})&+\beta(u_{n+2}-u_n-2\lambda)\label{K52}\\
&+(u_{n+1}-u_{n+2})(u_n-u_{n+1})(u_{n+2}-u_n-2\lambda).\notag
\end{align}
Note that once again $S_n$ has the property that as $\lambda \to 0$, $S_n \to I$, giving the periodic case. 

\vspace{.1cm}

To simply de-autonomize the lattice equation and the Lax pair, we let $\alpha_l=a+lh$ and $\beta_m=2mh$,
in which case the reduction, \eqref{reductioninQ}, becomes
\begin{equation}\label{dP1u}
u_{n+3} - u_n = \dfrac{a -hn -h}{u_{n+2}-u_{n+1}} + 2 \lambda,
\end{equation}
which we may transform to be a function of $y_n = u_{n+2}-u_{n+1}$, giving
\begin{equation}\label{dP1}
y_{n+1} + y_n + y_{n-1} = \dfrac{a -hn -h}{y_n} + 2 \lambda.
\end{equation}
This is a form of $\mathrm{d}$-$\mathrm{P}_{I}$ (see \cite{DPS}) and generalizes the reduction found in \cite{OvdKQ:reductions}. 
Furthermore, the method we present also gives us the Lax pair for this reduction. We specify our spectral parameter, given by
\eqref{spectralchoiceadd}, and construct $A_n(x)$ and $B_n(x)$ via their product representations, \eqref{prodnA} and \eqref{prodnB},
to give
\begin{align*}
A_n(x) =& \begin{pmatrix}
 u_{n+1}  & h+x+a - u_{n+1}(u_n+\lambda) \\
 1 & -u_n-\lambda 
\end{pmatrix} B_n(x),\\
B_n(x) =& \begin{pmatrix}
 u_{n+2}-\lambda  & x+a - u_{n+1}(u_{n+2}-\lambda) \\
 1 & \lambda-u_{n+2}
\end{pmatrix}
\begin{pmatrix}
 u_n  & hn+x - u_{n}(u_{n+2}-\lambda) \\
 1 & \lambda- u_{n+2}
\end{pmatrix}.
\end{align*}
Once again, we assume that $S_n$ is unknown, hence, we let $S_n$ be given by \eqref{arbitrarySn}. Then, using \eqref{calcS}, we find that $S_n$ is given by \eqref{H1Sn}. This gives us all the required elements of \eqref{compnonautadd}, which in turn, gives us \eqref{dP1u}. 

As in the modified Korteweg-de Vries reduction, it is possible to apply a transformation of the form of \eqref{gauge}, where 
\[
Z_{l,m} = \begin{pmatrix} 1 & w_{l,m} \\ 0 & 1 \end{pmatrix}
\]
to give the alternative Lax matrices, 
\begin{align*}
L_{l,m}&= 
\begin{pmatrix} 
 w_{l,m}-w_{l+1,m} & (w_{l,m}-w_{l+1,m})^2+\alpha_l-\gamma  \\
 1 & w_{l,m}-w_{l+1,m}
\end{pmatrix},\\
M_{l,m} &= 
\begin{pmatrix} 
 w_{l,m}-w_{l,m+1} & (w_{l,m}-w_{l,m+1})^2+\beta_m-\gamma  \\
 1 & w_{l,m}-w_{l,m+1}
\end{pmatrix},
\end{align*}
which have the desirable property of being expressed in terms of differences of the variables $w_{l,m}$. This means, these matrices admit a parameterisation in terms of Painlev\'e variables, $y_n =u_{n+2}-u_{n+1}$, 
\begin{align*}
A_n(x) =&  \begin{pmatrix} y_{n-1} - \lambda & a + x+ h+ (\lambda-y_{n-1})^2 \\ 
1 & y_{n-1} - \lambda \end{pmatrix}
B_n(x) ,\\
B_n(x) =&\begin{pmatrix} y_{n} - \lambda & a + x + (\lambda-y_{n})^2 \\ 
1 & y_{n} - \lambda \end{pmatrix}
\begin{pmatrix} \lambda - y_{n-1}-y_n  & x + nh + (\lambda-y_{n-1}-y_n)^2 \\ 
1 & \lambda -y_n - y_{n-1}\end{pmatrix}.
\end{align*}
We note that the transformation, $T$, applied to $y_n$ is trivial, just as in the previous section. This gives us that $S_n = I$, and the compatibility \eqref{compnonautmult} gives us \eqref{dP1}. This is not the first Lax pair known for equation \eqref{dP1}, as a $3\times 3$ Lax pair was derived in the work of Papageorgiou et al. \cite{Gramani:Isomonodromic}. We do not know whether a $2\times 2$ Lax pair, such as the one presented, is known or not. 

\vspace{.1cm}

We now turn to the fully non-autonomous twisted periodic reduction, where the $\alpha_l$ and $\beta_m$ variables are given by \eqref{H1fullvars}, with $b_0=0$. It was recently noted that the fully non-autonomous periodic reduction may be identified as a special case of the discrete analogue of the fourth Painlev\'e equation \cite{OvdKQ:reductions}. We expect this to be the case again.

As before, the evolution equations must take into account the way in which the $n \to n+1$ shift changes $l$ from an even number to an odd number, because the roles of $a_0$ and $a_1$ change every single iteration. The evolution equation, \eqref{reductioninQ}, in this case is given by 
\begin{equation} \label{eqPI}
u_{n+3}-u_n = \frac{- a_1+h n+ h}{u_{n+1}-u_{n+2}}+2 \lambda, \hspace{.7cm} a_0 \to a_1 - h, \hspace{.7cm} a_1 \to a_0 +h.
\end{equation}
Once again, it is not obvious that the mapping associated with the shift $n \to n+2$ is a two dimensional mapping. But we can find reduced variables
$y_n$ and $z_n$, by exploiting the Lax matrices for the equation, which are
\begin{align*}
A_n(x) =&  \begin{pmatrix} u_{n+1} & x+a_1-u_{n+1}(u_n+\lambda) \\ 
1 & -u_n-\lambda \end{pmatrix}
B_n(x) ,\\
B_n(x) =&\begin{pmatrix} u_{n+2}-\lambda & x+a_0-u_{n+1}(u_{n+2}-\lambda) \\ 
1 & -u_{n+1} \end{pmatrix} \begin{pmatrix} u_n  & x + nh - u_n(u_{n+2}-\lambda) \\ 
1 & \lambda-u_{n+2} \end{pmatrix}.
\end{align*}
The variables are explicitly given by
\begin{align*}
y_n =& -a_0+\left(u_n-u_{n+1}\right) \left(2 \lambda +u_n-u_{n+2}\right),\\
z_n =& \dfrac{a_0+y_n}{u_n-u_{n+1}}.
\end{align*}
These two functions of the lattice variables satisfy
\begin{subequations}
\begin{align}
y_{n+2} + y_n=& z_n(z_n - 2 \lambda)-a_0-a_1,\\
z_{n+2}z_n =& -\dfrac{(y_{n+2}+a_0)(y_{n+2} + a_1)}{y_{n+2} + h(n+2)},
\end{align}
\end{subequations}
which is a discrete version of the fourth Painlev\'e equation found in \cite{Gramani:coalescences, Quadratic}. This is a one parameter family of reductions that
generalizes the one presented in \cite{OvdKQ:reductions}.

On the other hand, the equation (\ref{eqPI}) is equivalent to asymmetric d-$\mathrm{P}_{\mathrm{I}}$, see \cite[Equation 3.33]{Quadratic}, where a relation to $d$-$\mathrm{P}_{IV}$ was obtained through a quadratic transformation. In fact, taking
\[
\alpha_l=a_1+a_2(-1)^l+hl,
\]
instead of (\ref{H1fullvars}), the equation then becomes
\[
u_{n+3} - u_n = \frac{h(n+1)-a_1+a_2(-1)^n}{u_{n+1}-u_{n+2}} + 2\lambda,
\]
or, if we let $y_n = u_{n+1} - u_n$, this becomes 
\[
y_{n+1} + y_n + y_{n-1} = \dfrac{hn - a_1 -a_2(-1)^n}{y_n} + 2\lambda 
\]
which is the most general form of $d$-$\mathrm{P}_{I}$ \cite{DPS}. In the autonomous limit, taking $h=0$, which would correspond to the ``fully autonomous case", the equation admits the following integral:
\[
K_{\eqref{QRT1}} + a_2(-1)^n(2u_{n+1}-u_n-u_{n+2}),
\]
where $K_{\eqref{QRT1}}$ is given in (\ref{K52}), taking $\alpha=a_1$, and $\beta=0$.

\vspace{.1cm}

Just as we did for \eqref{dmKdV}, we present a twist matrix for a
twist that is not homotopic to the identity twist, namely the twist
$T_2 : w \to \lambda - w$. The twist matrix associated with ($s_1,s_2)$-reductions
of \eqref{dpKdV} with a fixed Lax representation, \eqref{LaxH1}, is
\[
S_n = \begin{pmatrix} -1 & -\lambda \\ 0 & 1 \end{pmatrix},
\]
for a number of different choices of $s_1$ and $s_2$. 
This twist also yields a class of integrable mappings and their
Lax representations.

\section{Reductions of the lattice Schwarzian Korteweg-de Vries equation}\label{sec:dSKdV}

Periodic reductions of the lattice Schwarzian Korteweg-de Vries
equation (aka $Q1_{\delta = 0}$), given by \eqref{dSKdV}, have been the subject of a number of
studies \cite{Hay:Q1, Nijhoff:dSKdVP6, TKQ09}. Most recently, three of the
authors considered periodic reductions that gave rise to
$q$-$\mathrm{P}_{VI}$ and $q$-$\mathrm{P}(A_2^{(1)})$
\cite{OvdKQ:reductions}.

A Lax pair for equation  \eqref{dSKdV} is of the form \eqref{LaxQ} where
the Lax matrices are
\begin{align*}
L_{l,m} =& \begin{pmatrix}
1 & w_{l,m}-w_{l+1,m} \\[10pt]
\dfrac{\alpha }{\gamma (w_{l,m}- w_{l+1,m})} & 1
\end{pmatrix},\\
M_{l,m} =& \begin{pmatrix}
1 & w_{l,m}-w_{l,m+1} \\[10pt]
\dfrac{\beta }{\gamma(w_{l,m}- w_{l,m+1})} & 1
\end{pmatrix}.
\end{align*}
From our perspective, \eqref{dSKdV} is of particular interest, as it is
invariant under the full group of M\"obius transformations, denoted
$\mathrm{PGL}(2,\mathbb{C})$. We parameterise each M\"obius transformation
in terms of its fixed points, $\tau_1$ and $\tau_2$, and the eigenvalues, $\lambda_1$ and $\lambda_2$,
of a corresponding matrix, as follows:
\[
T(w) = \dfrac{(\lambda_1\tau_1- \lambda_2\tau_2)w - (\lambda_1 -
\lambda_2)\tau_1\tau_2}{(\lambda_1-\lambda_2)w +
\lambda_2\tau_1-\lambda_1\tau_2}.
\]
The reduced variables are given nicely in terms of $\tau_1$,
$\tau_2$, $\lambda_1$ and $\lambda_2$ as
\[
w_{l,m} \mapsto T^{l-m}u_{2m-l} = T^p u_n = \dfrac{(\lambda_1^p\tau_1-
\lambda_2^p\tau_2)u_n - (\lambda_1^p -
\lambda_2^p)\tau_1\tau_2}{(\lambda_1^p-\lambda_2^p)u_n +
\lambda_2^p\tau_1-\lambda_1^p\tau_2}.
\]
It will often be more notationally convenient to use the symbolic notation $T^p
u_n$ over the explicit expression for obvious reasons. In the autonomous case,
where $\alpha_l = \alpha$ and $\beta_m = \beta$ are constants, the
reduced equation may be expressed as
\begin{equation}\label{Q1Aut}
u_{n+3} =  \dfrac{\alpha  T^2u_n Tu_{n+1}-Tu_{n+2} ((\alpha
-\beta ) Tu_{n+1}+\beta  T^2u_{n})}{(\alpha -\beta )
T^2u_{n}-\alpha
  Tu_{n+2}+\beta  Tu_{n+1}}.
\end{equation}
We form the Lax pair in the usual manner, where \eqref{prodA} and
\eqref{prodB} give us the following representations for $A_n(\gamma)$
and $B_n(\gamma)$:
\begin{align*}
A_n(\gamma) =& \begin{pmatrix} 1 & u_{n+1} - T u_n \\[10pt]
\dfrac{\alpha}{\gamma(u_{n+1} - T u_n)} & 1 \end{pmatrix} B_n,\\
B_n(\gamma) =& \begin{pmatrix} 1 & T^{-1}u_{n+2} - u_{n+1} \\[10pt]
\dfrac{\alpha}{\gamma(T^{-1}u_{n+2} - u_{n+1})} & 1 \end{pmatrix}
 \begin{pmatrix} 1 & u_n - T^{-1}u_{n+2} \\[10pt] \dfrac{\beta}{\gamma(u_n
- T^{-1}u_{n+2})} & 1 \end{pmatrix}.
\end{align*}
The calculation of the twist matrix is algebraically more difficult
than in the previous cases, but essentially follows the same
logic. That is, we let $S_n$ be given by \eqref{arbitrarySn} and use
\eqref{calcS} at the various coefficients. The calculations are much
simpler if one assumes \eqref{Q1Aut}, but it is not necessary to do
so. It is also useful to compare the iterates of the entries of $S_n$
with the calculated values for $S_{n+1}$. This gives us our third
non-trivial twist matrix, associated with the M\"obius transformation,
given by
\begin{equation}\label{twistQ1}
S_n = \begin{pmatrix}
\dfrac{\lambda_1\lambda_2 (\tau_1-\tau_2)}{\lambda_1(u_n-\tau_2)
-\lambda_2(u_n-\tau_1)} & 0\\[10pt]
\dfrac{\lambda_1-\lambda_2}{\tau_1-\tau_2} &
\dfrac{\lambda_1(u_n-\tau_2) - \lambda_2(u_n-\tau_1)}{\tau_1 - \tau_2}
\end{pmatrix}.
\end{equation}
The coefficient of $\gamma^{-1}$ in the trace of the twisted monodromy matrix
provides the following integral for equation \eqref{Q1Aut}:
\begin{align*}
K_{\eqref{Q1Aut}}
=&\frac{\alpha(Tu_{n+1}-T^2u_{n})(\lambda_{1}(\tau_{2}-u_{n+2})-\lambda_{2}(\tau_{1}-u_{n+2}))}{\lambda_{1}\lambda_{2}(\tau_{1}-\tau_{2})(Tu_{n+1}-u_{n+2})}\\
&+\frac{\beta(u_{n+2}-T^2u_{n})(\lambda_{1}(Tu_{n}-\tau_{2})-\lambda_{2}(Tu_{n}-\tau_{1}))}{\lambda_{1}\lambda_{2}(\tau_{1}-\tau_{2})(Tu_{n}-u_{n+2})}\\
&+\frac{\alpha(\tau_{1}-\tau_{2})(Tu_{n}T^2u_{n}-(Tu_{n+1})^2+u_{n+2}(2Tu_{n+1}-Tu_{n}-T^2u_{n}))}{(\lambda_{1}(Tu_{n}-\tau_{2})-\lambda_{2}(Tu_{n}-\tau_{1}))(Tu_{n+1}-T^2u_{n})(Tu_{n+1}-u_{n+2})}.
\end{align*}
We have determined the reduced variables to be
\begin{align*}
y_n &= \dfrac{\beta\left(Tu_n-u_n\right) \left(T^{-1}u_{n+2}-u_{n+1}\right)}{\alpha
  \left(Tu_n-u_{n+1}\right) \left(T^{-1}u_{n+2}-u_n\right)},\\
z_n &= \dfrac{\lambda _1 \lambda _2 \left(\tau _1-\tau _2\right)
(\alpha y_n -1) (T^{-1}u_{n+2}-u_n)}{(Tu_n-T^{-1}u_{n+2})
  \left(\lambda _2 \left(\tau _1-u_n\right)+\lambda _1 \left(u_n-\tau
_2\right)\right)},
\end{align*}
and hence we obtain the equation
\begin{subequations} \label{Q1AutR}
\begin{align}
y_{n+1}y_n =& \dfrac{\beta\left(z_n-\lambda_1\right)\left( z_n
-\lambda_2\right)}{\alpha \lambda_1 \lambda_2},\\
z_{n+1}z_n =& (1-y_{n+1})\lambda_1\lambda_2,
\end{align}
\end{subequations}
which is of QRT type and admits the integral
\[
K_{\ref{Q1AutR}}=\alpha \left( \frac{ y_n-1}{z_n}-\frac{z_n}{\lambda_1\lambda_2}\right)
+\beta\left(\frac{\left(z_n-\lambda _1\right) \left(z_n-\lambda _2\right)}{\lambda _1 \lambda _2 y_n
  z_n}-\frac{1}{z_n}\right).
\]

\hspace{.1cm}

Let us jump right to the fully non-autonomous reduction, where the
variables $\alpha_l$ and $\beta_m$ are given by \eqref{H3Q1fullvars},
with $b_0=1$. If we assume $l$ (and hence $n$) is even, then the evolution equation is
given by
\begin{align}
&a_1 \to q a_0 ,\hspace{1cm} a_0 \to \dfrac{a_1}{q},\label{Q1nonautred}\\
&u_{n+3}=\frac{a_1 T u_{n+1} \left(T u_{n+2}-T^2 u_n\right)+ q^{n+2}
  T u_{n+2} \left(T^2 u_n-T u_{n+1}\right)}{a_1
  \left(T u_{n+2}-T^2u_n\right)+ q^{n+2}
  \left(T^2u_n-Tu_{n+1}\right)}\nonumber.
\end{align}
The Lax matrices are given by \eqref{prodnA} and \eqref{prodnB},
\begin{align*}
A_n(x) =& \begin{pmatrix} 1 & u_{n+1} - T u_n \\[10pt] \dfrac{xa_1}{u_{n+1}
- T u_n} & 1 \end{pmatrix} B_n(x), \\
B_n(x) =& \begin{pmatrix} 1 & T^{-1}u_{n+2} - u_{n+1} \\[10pt] \dfrac{x
a_0}{T^{-1}u_{n+2} - u_{n+1}} & 1 \end{pmatrix} 
\begin{pmatrix}
1 & u_n - T^{-1}u_{n+2} \\[10pt] \dfrac{xq^n}{u_n - T^{-1}u_{n+2}} & 1
\end{pmatrix}.
\end{align*}
We use \eqref{calcS} to deduce that $S_n$ is again given by
\eqref{twistQ1}. Using the compatibility, \eqref{compnonautmult}, we
readily find \eqref{Q1nonautred}. Once again, the task remains to find
a second order system from this equation. We choose a similar
combination of lattice variables as before, by letting
\begin{align*}
y_n &= \dfrac{(Tu_n-u_{n}) (T^{-1}u_{n+2}-u_{n+1})}{a_0
(Tu_{n}-u_{n+1}) (T^{-1}u_{n+2}-u_n)},\\
z_n &= \dfrac{(T^{-1}u_{n+2}-Tu_n) \left(\lambda _2 \left(\tau
_1-u_{n}\right)+\lambda _1 \left(u_{n}-\tau
_2\right)\right)}{\left(\tau
  _1-\tau _2\right) (T^{-1}u_{n+2}-u_{n})}.
\end{align*}
Under this change of variables, we obtain another version of the system obtained in \cite{Quadratic}, which generalizes \eqref{H3nonautred},
\begin{subequations}
\begin{align}
y_{n+2} y_n =& \dfrac{(z_n-\lambda_1)(z_n-\lambda_2)}{\lambda_1\lambda_2 a_0 a_1},\\
z_{n+2} z_n =& \dfrac{\lambda_1\lambda_2\left(a_0 y_{n+2}-1\right)
\left(a_1 y_{n+2}-1\right)}{1- q^{n+2} y_{n+2}},
\end{align}
\end{subequations}
modulo a certain scaling of variables. It is interesting to note that as a system admitting singularity
confinement, the critical values of $z_n$ depend explicitly on the eigenvalues of the twist. 

\section{($2,2$)-reduction, and $q$-$\mathrm{P}_{VI}$} \label{22qpvi}
Three of the authors have presented two versions of $q$-$\mathrm{P}_{VI}$, from
\eqref{dmKdV} in \cite{Ormerod:qP6} and from \eqref{dSKdV} in
\cite{OvdKQ:reductions}. Both of these reductions were subcases of the
system described in the work of Jimbo and Sakai \cite{Sakai:qP6}; the
version in \cite{OvdKQ:reductions} appeared with an interesting
biquadratic constraint, which was similar to the work of Yamada
\cite{Yamada:LaxqEs} but not present in \cite{Sakai:qP6}, while the
version in \cite{Ormerod:qP6} is a subcase of the version in
\cite{OvdKQ:reductions}. Here we will present the fully non-autonomous $(2,2)$-reduction
of \eqref{dSKdV}, which we identify with the full parameter unconstrained version of the
$q$-analogue of the sixth Painlev\'e equation as it appears in \cite{Sakai:qP6}.

We start by specifying new $n$ and $p$ variables, which we assign to be
\[
n= m-l, \hspace{1cm} p = \left\lfloor \dfrac{l}{2} \right\rfloor,
\]
where $\left\lfloor x \right\rfloor$ rounds $x$ down to the nearest
integer. In this way, we label the variables $w_{l,m}$ so that
\begin{equation} \label{labelling22}
w_{l,m} \mapsto \left\{ \begin{array}{l p{4cm}} T^p u_n & if $l$ is
even,\\T^p v_n & if $l$ is odd. \end{array} \right.
\end{equation}
This labeling is depicted in Figure \ref{fig:labelqP6}.

\begin{figure}[!ht]
\begin{tikzpicture}[scale=1.5]
\draw[black!30] (-1.5,-.5) grid (3.5,3.5);
\draw[blue,thick] (-1,-.5)--(-1,0) -- (0,0) -- (0,1) -- (1,1)--
(1,2)-- (2,2)-- (2,3)--(3,3)--(3,3.5);
\begin{scope}[xshift=.3cm,yshift=-.2cm]
\node at (-.1,0) {$T^pu_n$};
\node at (0,1) {$T^pu_{n+1}$};
\node at (0,2) {$T^pu_{n+2}$};
\node at (.9,1) {$T^pv_{n}$};
\node at (1,2) {$T^pv_{n+1}$};
\node at (1,3) {$T^pv_{n+2}$};
\node at (2,2) {$T^{p+1}u_{n}$};
\node at (2.1,3) {$T^{p+1} u_{n+1}$};
\node at (3,3) {$T^{p+1} v_{n}$};
\end{scope}
\filldraw[red] (0,2) circle (.07);
\filldraw[red] (1,3) circle (.07);
\end{tikzpicture}
\caption{Labeling of variables for the (2,2)-reduction of the lattice
with respect to \eqref{labelling22}. \label{fig:labelqP6}}
\end{figure}

In order for this system to be consistent, we require
\[
\dfrac{\alpha_{l+2}}{\beta_{m+2}} = \dfrac{\alpha_{l}}{\beta_{m}},
\]
which we solve by letting
\[
\alpha_l = \left\{ \begin{array}{c p{2cm}}
a_0 q^l & if $l$ is even,\\
a_1 q^l & if $l$ is odd,\end{array}\right. \hspace{1cm}
\beta_m = \left\{ \begin{array}{c p{2cm}}
b_0 q^m & if $m$ is even,\\
b_1 q^m & if $m$ is odd.\end{array}\right.
\]
We now pick a spectral variable, $x= q^l/\gamma$, in which we have the
system of linear equations
\begin{align*}
TY_{n}(q^2 x) &= A_n(x) Y_n(x),\\
Y_{n}(x) &= B_n(x) Y_n(x),
\end{align*}
where the spectral matrix, $A_n(x)$, governs an operation that is
equivalent to the shift $(l,m) \to (l+2,m+2)$ and the deformation
matrix, $B_n(x)$, governs an operation that is equivalent to the shift
$(l,m) \to (l,m+1)$. This gives us a linear system with Lax matrices
\begin{align*}
A_n(x) &\mapsfrom L_{l+1,m+2}M_{l+1,m+1}L_{l,m+1}M_{l,m},\\
B_n(x) &\mapsfrom M_{l,m}.
\end{align*}
explicitly given by
\begin{align*}
A_n(x) = &\begin{pmatrix} 1 & v_{n+1}-T u_n \\ \frac{x a_1}{v_{n+1}-T
u_n} & 1 \end{pmatrix} \begin{pmatrix} 1 & v_n-v_{n+1} \\ \frac{q^n x
b_1}{v_n-v_{n+1}} & 1\end{pmatrix}\\
& \times \begin{pmatrix}1 & u_{n+1}-v_n \\ \frac{x a_0}{u_{n+1}-v_n} & 1
\end{pmatrix} B_n(x), \\
B_n(x) = & \begin{pmatrix}1 & u_n-u_{n+1} \\ \frac{q^n x
b_0}{u_n-u_{n+1}} &1\end{pmatrix}.
\end{align*}
The twist matrix $S_n$ is the same as in the (2,1)-reduction, given by
\eqref{twistQ1}, and the compatibility condition
\[
S_{n+1}^{-1} A_{n+1}(x)B_n(x) - B_n(q^2x) S_n^{-1} A_n(x) =0
\]
gives the system that fixes the $a_0$ and $a_1$, and induces the transformation
\begin{align*}
&b_0 \to \dfrac{b_1}{q}, \hspace{1cm} b_1 \to q b_0,\\
u_{n+2} &= \dfrac{a_0 u_{n+1} \left(v_n-v_{n+1}\right)+b_1 q^n v_{n+1}
\left(u_{n+1}-v_n\right)}{a_0 \left(v_n-v_{n+1}\right)+b_1 q^n
  \left(u_{n+1}-v_n\right)},\\
v_{n+2} &= \dfrac{a_1 v_{n+1} \left(T u_n-T u_{n+1}\right)+b_0 q^{n+2} Tu_{n+1}
  \left(v_{n+1}-T u_n\right)}{a_1 \left(T u_n-T u_{n+1}\right)+b_0 q^{n+2}
  \left(v_{n+1}-T u_n\right)}.
\end{align*}
This system possesses a $2$-integral\footnote{A $2$-integral is invariant under the second iterate of the map \cite{HBQC}.}, which we label $\kappa$, given by
\[
\kappa = \dfrac{b_0 q^n \left(u_{n+1}-v_n\right) \left(\lambda _2
\left(\tau _1-u_n\right)+\lambda _1 \left(u_n-\tau _2\right)\right)
  \left(Tu_n-v_{n+1}\right)}{a_1 \lambda _1 \left(\tau _2-\tau
_1\right) \left(u_n-u_{n+1}\right)
  \left(v_n-v_{n+1}\right)}.
\]
The $A_n(x)$ may be identified with the parameterisation of the
spectral matrix of Jimbo and Sakai (see \cite{Sakai:qP6}) with
variables $y_n$ and $z_n$ specified by
\begin{align*}
y_n&= \dfrac{b_0 q^n \left(v_n-u_{n+1}\right) \left(\lambda _2
\left(\tau _1-u_n\right)+\lambda _1 \left(u_n-\tau _2\right)\right)
  \left(T u_n-v_{n+1}\right)}{a_1 \lambda _1 \left(\tau _1-\tau
_2\right) \left(u_n-u_{n+1}\right)
  \left(v_n-v_{n+1}\right)},\\
z_n &=\frac{\left(v_{n+1}-v_{n}\right) \left(T u_n-u_n\right)
\left(u_{n+1}-v_n\right)}{a_0 \left(u_n-u_{n+1}\right)
  \left(v_n-v_{n+1}\right) \left(Tu_n-v_n\right)+b_1 q^n
\left(u_n-v_n\right) \left(u_{n+1}-v_n\right)
  \left(Tu_n-v_{n+1}\right)}.
\end{align*}
Under this change of variables, the system takes the form
\begin{subequations}
\begin{align}
y_{n+2} y_n &= \dfrac{\kappa (z_n-
\lambda_1)(z_n-\lambda_2)}{(a_1\kappa
z_n-q^n\lambda_2b_0)(a_0z_n-q^{n+2}\kappa \lambda_1b_1)},\\
z_{n+2}z_n &=\frac{\lambda _1 \lambda _2 \left(b_0 q^{n+2}
y_{n+2}-1\right) \left(b_1 q^{n+2} y_{n+2}-1\right)}{\left(a_0
  y_{n+2}-1\right) \left(a_1 y_{n+2}-1\right)}.
\end{align}
\end{subequations}
This is the $q$-analogue of the sixth Painlev\'e equation \cite{Sakai:qP6}.

\section{General $(s_1,s_2)$-reduction} \label{s1s2}
We have specified several cases of $(2,1)$-reductions and a single
case of a $(2,2)$-reduction, however, this theory generalizes to an
arbitrary $(s_1,s_2)$-reduction. Following \cite{OvdKQ:reductions}, we let $s_1
= ag$ and $s_2 = bg$, with gcd($a,b$)=1. Then we specify two integers, $c$ and $d$, by
\[
\det \begin{pmatrix} a & b \\ c & d \end{pmatrix} =1.
\]
From this we define the variables
\begin{equation} \label{nk}
n =n(l,m)= \det\begin{pmatrix} a & b \\ l & m \end{pmatrix},
\hspace{1cm} k =k(l,m)= \det \begin{pmatrix} l & m \\ c & d
\end{pmatrix} \mod g,
\end{equation}
and
\begin{equation} \label{p}
p = p(l,m)= \left\lfloor  \dfrac{1}{g} \det \begin{pmatrix} l & m \\ c
& d \end{pmatrix} \right\rfloor.
\end{equation}
Now we perform the reduction in accordance with the rule
\begin{equation}
w_{l,m} \mapsto T^p u_n^k.
\end{equation}
We note that the $p$ variable is the power of the transformation, $T$,
whereas the $k$ is a superscript. The general ($s_1,s_2$)-reduction of \eqref{Q} is given
by the system of $g$ equations:
\begin{equation}
Q(T^{p} u_n^k, T^{\tilde{p}}u_{n-b}^{k+d}, T^{\hat{p}}u_{n+a}^{k-c},
T^{\hat{\tilde{p}}}u_{n+a-b}^{k-c+d};\alpha, \beta) = 0, \hspace{1cm}
k = 0,1, \ldots, g-1,
\end{equation}
where the superscripts are interpreted modulo $g$ and $\tilde{p} =
p(l+1,m)$ and $\hat{p} = p(l,m+1)$ are just the expressions for the
$p$'s shifted in the $l$ and $m$ directions respectively. This choice of
labels and powers of $T$ ensures that any two ways of calculating an iterate, $u_n^k$, 
coincide due to the invariance of $Q$ under the action of the twist, $T$.

We construct operators that govern the shifts $(l,m) \to (l+s_1,m+s_2)$ and $(l,m) \to (l+c,m+d)$, which
have the effect
\begin{subequations}\label{Laxaut}
\begin{align}
\label{LaxLn} T\Psi_n &= A_n \Psi_n,\\
\label{LaxMn}\Psi_{n+1} &= B_n \Psi_n,
\end{align}
\end{subequations}
in which the matrices, $A_n$ and $B_n$, can be specified by
\begin{subequations}
\begin{align}
\label{prodAl}A_n &\mapsfrom \prod_{j=0}^{s_2-1}
M_{l+s_1,m+j}\prod_{i=0}^{s_1-1} L_{l+i,m},\\
\label{prodBl}B_n &\mapsfrom \prod_{j=0}^{d-1}
M_{l+c,m+j}\prod_{i=0}^{c-1} L_{l+i,m},
\end{align}
\end{subequations}
and $n$ is given by \eqref{nk}, see also \cite{OR1,OR2}, where Lax matrices $A_n$
and $B_n$ are given in terms of a product along so called standard staircases.
The determining equation, \eqref{twistmatrix}, that defines the twist matrix is
also a valid ansatz for the general $(s_1,s_2)$-reduction.

A list of possible (M\"obius) twists for the equations
in the ABS-list appears in Table
\ref{tableTlist}.

\begin{table}[!ht]
\begin{tabular}{l l} \toprule
ABS & Point Symmetries \\ \toprule
$H1$ & $T_1:w \to w+\lambda$,\,\, $T_2 : w \to \mu-w$,\\
$H3_{\delta=0}$ & $T_1:w \to \lambda w,\,\, T_2 : w \to \dfrac{\mu}{w}$,\\
$H3_{\delta\neq0}$ & $T_1:w \to -w$,\\
$Q1_{\delta\neq0}$ & $T_1:w \to w+\lambda$,\,\, $T_2 : w \to \mu-w$,\\
$Q1_{\delta= 0}$ & $T_1:w \to \dfrac{(\lambda_1\tau_1- \lambda_2\tau_2)w
- (\lambda_1 - \lambda_2)\tau_1\tau_2}{(\lambda_1-\lambda_2)w +
\lambda_2\tau_1-\lambda_1\tau_2}$,\\
$Q3_{\delta=0}$ & $T_1 : w \to \lambda w$, \,\, $T_2 : w \to \mu/w$,\\
$Q3_{\delta\neq 0}$ & $T_1 : w \to -w$,\\
$Q4$ & $T_1 : w \to -w$,\,\, $T_2 : w \to 1/w$,\\
$A1_{\delta=0}$ & $T_1 : w \to \lambda w$,\,\, $T_2 : w \to \mu/w$,\\
$A1_{\delta \neq 0}$ & $T_1 : w \to -w$,\\ \bottomrule
\end{tabular}
\caption{A list of the M\"obius point symmetries of the lattice
equations that appear in the ABS list \cite{ABS:ListI,ABS:ListII}. For Q4 we used the version given in \cite{Hie05}.}
\label{tableTlist}
\end{table}

Let us conclude by mentioning twisted reductions for non-autonomous multiplicative equations, i.e., those for which $Q$ and the Lax-matrices depend on $\alpha/\beta$ only. Under this assumption\footnote{The case for additive type twisted reductions can
be formulated analogously to what is presented here.} the reduction is consistent, provided
$
\alpha_{l+s_1}/\beta_{m+s_2}= \alpha_l/\beta_m.
$
By separation of variables this gives
\[
\dfrac{\alpha_{l+s_1}}{\alpha_{l}} = \dfrac{\beta_{m+s_2}}{\beta_m} := q^{abg},
\]
which is solved by
\begin{equation}
\alpha_l =  a_{l\,\, \mathrm{mod}\, s_1} q^{bl}, \hspace{2cm} \beta_m
=  b_{m\,\, \mathrm{mod}\, s_2} q^{am}.
\end{equation}
A simple choice of spectral variable is $x = q^l$, in which the product
representations of $A_n$ and $B_n$, given above by \eqref{prodAl} and
\eqref{prodBl}, depend on $x$, giving $A_n(x)$ and $B_n(x)$. These
matrices define a linear system
\begin{subequations}
\begin{align}
T Y_n(q^{abg} x) = A_n(x) Y_n(x),\\
Y_{n+1}(q^{cb} x) = B_n(x) Y_n(x),
\end{align}
\end{subequations}
along with the definition of the twist matrix,
\eqref{nonauttwistmatrix}, gives the compatibility
\begin{equation}
S_{n+1}^{-1} A_{n+1}(q^{cb}x)B_n(x) - B_{n}(q^{abg}x)S_n^{-1}A_n(x) = 0.
\end{equation}
This compatibility is equivalent to the system of $g$ equations that
define the non-autonomous reductions,
\begin{equation}
Q(T^{p} u_n^k, T^{\tilde{p}}u_{n-b}^{k+d}, T^{\hat{p}}u_{n+a}^{k-c},
T^{\hat{\tilde{p}}}u_{n+a-b}^{k-c+d};\alpha_l/\beta_m) = 0,
\hspace{1cm} k = 0,1, \ldots, g-1,
\end{equation}
where we should recall that $\alpha_l/\beta_m$, as a function of $n$, is
\[
\dfrac{\alpha_l}{\beta_m} = \dfrac{a_{l\, \mathrm{mod}\, s_1}}{b_{m\,
\mathrm{mod}\, s_2}} q^{-n}.
\]
This provides a Lax representation for the twisted $(s_1,s_2)$-reduction with general $s_1$ and $s_2$.

\section{Conclusions}
We have presented a generalisation of periodic reductions, that would appear to be new. Applying this to integrable equations, the resulting reductions possess Lax representations. This method can be used to obtain many additional integrable mappings. This can be done either by considering other reductions, or by starting from other integrable equations on quads (both of ABS type or non-ABS type), or from (multi-component) equations on other stencils. The method proposed in this paper seems analogous to Sklyanin's method for generalising periodic boundary conditions for integrable quantum systems \cite{S88}. Finally we note that twisted reductions may also apply to non-integrable equations (although in that case there will be no Lax representations).

\section*{Acknowledgments}
We are grateful to Dr V. Mangazeev for a question posed at the ANZAMP Inaugural Meeting in Lorne, Dec 2012, that let to this paper being written. We would also like to thank Dr. B. Grammaticos for some references on discrete Painlev\'e equations. This research is supported by Australian Research Council Discovery Grants \#DP110100077 and \#DP140100383.


\begin{thebibliography}{10}

\bibitem{AblowitzClarkson:Solitons}
M.~J. Ablowitz and P.~A. Clarkson.
\newblock  Solitons, nonlinear evolution equations and inverse scattering,
  volume 149 of London Mathematical Society Lecture Note Series.
\newblock Cambridge University Press, Cambridge, 1991.

\bibitem{AblowitzSegur}
M.~J. Ablowitz and H. Segur.
\newblock Solitons and the inverse scattering transform
\newblock SIAM, Philadelphia, 1981.

\bibitem{ABS:ListI}
V.~E. Adler, A.~I. Bobenko, and Y.~B. Suris.
\newblock Classification of integrable equations on quad-graphs. {T}he
  consistency approach.
\newblock {\em Comm. Math. Phys.}, 233(3):513--543, 2003.

\bibitem{ABS:ListII}
V.~E. Adler, A.~I. Bobenko, and Y.~B. Suris.
\newblock Discrete nonlinear hyperbolic equations: classification of integrable
  cases.
\newblock {\em Funct. Anal. Appl.}, 43(1):3--17, 2009.

\bibitem{BC74}
G.~W. Bluman, and J.~D. Cole.
\newblock Similarity methods for differential equations.
\newblock {\em Appl. Math. Sci.}, 13, Springer-Verlag, New York, 1974.

\bibitem{Butler:scatteringKdV}
S. Butler and N. Joshi.
\newblock An inverse scattering transform for the lattice potential KdV
  equation.
\newblock {\em Inverse Problems}, 26(11):115012, 2010.

\bibitem{DISPI}
K.~M. Case, and M. Kac.
\newblock A discrete version of the inverse scattering problem.
\newblock {\em J. Math. Phys}, {\bf 14} (5), 594--603, 1973.

\bibitem{DISPII}
K.~M. Case.
\newblock On discrete inverse scattering problems. II.
\newblock {\em J. Math. Phys}, {\bf 14} (5), 916--920, 1973.

\bibitem{Jimbo:discretesolitonIII}
E. Date, M. Jimbo, and T. Miwa.
\newblock Method for generating discrete soliton equations. {III}.
\newblock {\em J. Phys. Soc. Japan}, 52(2):388--393, 1983.

\bibitem{Jimbo:discretesolitonII}
E. Date, M. Jimbo, and T. Miwa.
\newblock Method for generating discrete soliton equations. {IV}, {V}.
\newblock {\em J. Phys. Soc. Japan}, 52(3):761--765, 766--771, 1983.

\bibitem{Jimbo:discretesolitonI}
E. Date, M. Jimbo, and T. Miwa.
\newblock Method for generating discrete soliton equations. {I}, {II}.
\newblock {\em J. Phys. Soc. Japan}, 51(12):4116--4124, 4125--4131, 1982.

\bibitem{DuVal}
P. Du Val. 
\newblock Homographies, quaternions, and rotations.
\newblock {\em Clarendon Press}, Oxford 1964.

\bibitem{FlaschkaNewell}
H. Flaschka and A.~C. Newell.
\newblock Monodromy- and spectrum-preserving deformations. {I}.
\newblock {\em Comm. Math. Phys.}, 76(1):65--116, 1980.

\bibitem{Hunting}
B.~Grammaticos and A.~Ramani. 
\newblock The hunting for the discrete Painlev\'e equations. 
\newblock {\em Reg. and Chaot. Dyn.}, 5(1), 53-66, 2000.

\bibitem{Gramani:Reductions}
B.~Grammaticos, A.~Ramani, J.~Satsuma, R.~Willox, and A.~S. Carstea.
\newblock Reductions of integrable lattices.
\newblock {\em J. Nonlinear Math. Phys.}, 12(suppl. 1):363--371, 2005.

\bibitem{HBQC}
F. Haggar, G.B. Byrnes, G.R.W. Quispel, and H.W. Capel.
\newblock $k$-integrals and $k$-Lie symmetries in discrete dynamical systems.
\newblock {\em Physica} 233A: 379-394, 1996.

\bibitem{Hay}
M. Hay, J. Hietarinta, N. Joshi, and F.~W. Nijhoff.
\newblock A {L}ax pair for a lattice modified {K}d{V} equation, reductions to
  {$q$}-{P}ainlev\'e equations and associated {L}ax pairs.
\newblock {\em J. Phys. A}, 40(2):F61--F73, 2007.

\bibitem{HHS13}
M. Hay, P. Howes, and Y. Shi.
\newblock A systematic approach to reduction of type-Q ABS equations.
\newblock {\em arXiv:}1307.3390v1 [nlin.SI].

\bibitem{Hay:Q1}
M. Hay, K. Kajiwara, and T. Masuda.
\newblock Bilinearisation and special solutions to the discrete {S}chwarzian
  {K}d{V} equation.
\newblock {\em J. Math-for-Ind.}, 3A:53--62, 2011.

\bibitem{HKQT13}
A.N.W. Hone, P.H. van der Kamp, G.R.W. Quispel, and D.T. Tran.
\newblock Integrability of reductions of the discrete KdV and potential KdV equations.
\newblock {\em Proc R Soc A} 469: 20120747.

\bibitem{Hie05} 
J. Hietarinta. 
\newblock Searching for CAC-maps.
\newblock {\em J. Nonlin. Math. Phys.}, 12 (2): 223--230, 2005.

\bibitem{Hirota:DKdV}
R. Hirota.
\newblock Nonlinear partial difference equations. {I}. {A} difference analogue
  of the {K}orteweg-de {V}ries equation.
\newblock {\em J. Phys. Soc. Japan}, 43(4):1424--1433, 1977.

\bibitem{Hirota:DtToda}
R. Hirota.
\newblock Nonlinear partial difference equations. {II}. {D}iscrete-time {T}oda
  equation.
\newblock {\em J. Phys. Soc. Japan}, 43(6):2074--2078, 1977.

\bibitem{Hirota:dSG}
R. Hirota.
\newblock Nonlinear partial difference equations. {III}. {D}iscrete
  sine-{G}ordon equation.
\newblock {\em J. Phys. Soc. Japan}, 43(6):2079--2086, 1977.

\bibitem{Sakai:qP6}
M. Jimbo and H. Sakai.
\newblock A {$q$}-analog of the sixth {P}ainlev\'e equation.
\newblock {\em Lett. Math. Phys.}, 38(2):145--154, 1996.

\bibitem{AsymmetricdPs}
M. D. Kruskal, K. M. Tamizhmani, B. Grammaticos, and A. Ramani. 
\newblock Asymmetric discrete Painlev\'e equations. 
\newblock {\em Regular and Chaotic Dynamics}, 5(3), 273-280, 2000.

\bibitem{LeviW}
D. Levi and P. Winternitz.
\newblock Symmetries and conditional symmetries of differential-difference equations.
\newblock {\em J. Math. Phys.}, 34 (8), 3713--3730, 1993. 

\bibitem{Nij02} F.W. Nijhoff, \textit{Lax pair for the Adler (lattice
    Krichever-Novikov) system}, Phys. Lett. {\bf 297A} (2002), 49--58.

\bibitem{SimilarityReds}
F.~W. Nijhoff and V.~G. Papageorgiou.
\newblock Similarity reductions of integrable lattices and discrete analogues
  of the {P}ainlev\'e {${\rm II}$} equation.
\newblock {\em Phys. Lett. A}, 153(6-7):337--344, 1991.

\bibitem{Nijhoff:dSKdV}
F.~W. Nijhoff, G.~R.~W. Quispel, and H.~W. Capel.
\newblock Direct linearisation of nonlinear difference-difference equations.
\newblock {\em Phys. Lett. A}, 97(4):125--128, 1983.

\bibitem{Nijhoff:Linearisation}
F.~W. Nijhoff, G.~R.~W. Quispel, and H.~W. Capel.
\newblock Linearisation of nonlinear differential-difference equations.
\newblock {\em Phys. Lett. A}, 95(6):273--276, 1983.

\bibitem{Nijhoff:dSKdVP6}
F.~W. Nijhoff, A.~Ramani, B.~Grammaticos, and Y.~Ohta.
\newblock On discrete {P}ainlev\'e equations associated with the lattice
  {K}d{V} systems and the {P}ainlev\'e {VI} equation.
\newblock {\em Stud. Appl. Math.}, 106(3):261--314, 2001.

\bibitem{NijWal}
 F.~W. Nijhoff and A.~J. Walker.
\newblock The discrete and continuous Painlev\'e VI hierarchy and the Garnier systems.
\newblock Glasgow Math. J. 43A 109--23, 2001.

\bibitem{Olver:LieDiff}
P.~J. Olver.
\newblock {\em Applications of {L}ie groups to differential equations}, volume
  107 of {\em Graduate Texts in Mathematics}.
\newblock Springer-Verlag, New York, 1986.

\bibitem{OR1}
O. Rojas, P.~H. van~der Kamp, and G.~R.~W. Quispel.
\newblock Lax representation for integrable O$\Delta$Es,
\newblock {\em proceedings `Symmetry and Perturbation Theory 2007'}, 271--272, 2008.

\bibitem{OR2}
O. Rojas, 
\newblock From discrete integrable systems to cellular automata. 
\newblock PhD thesis, La Trobe University, 2009.

\bibitem{OvdKQ:reductions}
C.~M. Ormerod, P.~H. van~der Kamp, and G.~R.~W. Quispel.
\newblock Discrete Painlev\'e equations and their Lax pairs as reductions of
  integrable lattice equations.
\newblock {\em J. Phys. A} , 46(9):095204, 2013.

\bibitem{Ormerod:qP6}
C.~M. Ormerod.
\newblock Reductions of lattice m{K}d{V} to {$q$}-{${\rm P}_{\rm VI}$}.
\newblock {\em Phys. Lett. A}, 376(45):2855--2859, 2012.

\bibitem{dKdVreds}
V.~G. Papageorgiou, F.~W. Nijhoff, and H.~W. Capel.
\newblock Integrable mappings and nonlinear integrable lattice equations.
\newblock {\em Phys. Lett. A}, 147(2-3):106--114, 1990.

\bibitem{Gramani:Isomonodromic}
V.~G. Papageorgiou, F.~W. Nijhoff, B. Grammaticos, and A. Ramani.
\newblock Isomonodromic deformation problems for discrete analogues of
  {P}ainlev\'e equations.
\newblock {\em Phys. Lett. A}, 164(1):57--64, 1992.

\bibitem{PWZ:I}
J. Patera, P. Winternitz and H. Zassenhaus.
\newblock Continuous subgroups of the fundamental groups of physics. I. General method and the Poincar\'e group. 
\newblock {\em J. Math. Phys.}, 16, 1597--1614, 1975. 

\bibitem{PWZ:II}
J. Patera, P. Winternitz and H. Zassenhaus.
\newblock Continuous subgroups of the fundamental groups of physics. II. The similitude group.
\newblock {\em J. Math. Phys.}, 16, 1615--1624, 1975. 

\bibitem{QCS}
G.R.W. Quispel, H.W. Capel and R. Sahadevan,
\newblock Continuous symmetries of differential-difference equations: the Kac-van Moerbeke equation and Painlev\'e reduction.
\newblock {\em Phys. Lett.}, 170A, 379--383, 1992.

\bibitem{Quispel:Intreds}
G.~R.~W. Quispel, H.~W. Capel, V.~G. Papageorgiou, and F.~W. Nijhoff.
\newblock Integrable mappings derived from soliton equations.
\newblock {\em Physica A}, 173(1-2):243--266, 1991.

\bibitem{QNCvdL:NSG}
G.~R.~W. Quispel, F.~W. Nijhoff, H.~W. Capel, and J.~van~der Linden.
\newblock Linear integral equations and nonlinear difference-difference
  equations.
\newblock {\em Phys. A}, 125(2-3):344--380, 1984.

\bibitem{QRT1}
G.~R.~W. Quispel, J.~A.~G. Roberts, and C.~J. Thompson.
\newblock Integrable mappings and soliton equations.
\newblock {\em Phys. Lett. A}, 126(7):419--421, 1988.

\bibitem{QRT2}
G.~R.~W. Quispel, J.~A.~G. Roberts, and C.~J. Thompson.
\newblock Integrable mappings and soliton equations. {II}.
\newblock {\em Phys. D}, 34(1-2):183--192, 1989.

\bibitem{Gramani:Q4Ell}
A. Ramani, A.~S. Carstea and B. Grammaticos.
\newblock On the non-autonomous form of the {Q}4 mapping and its
relation to elliptic {P}ainlev\'e equations.
\newblock {\em J. Phys. A}  42 (32), 2009.

\bibitem{ramani2002autonomous}
A. Ramani, A.~S. Carstea, B. Grammaticos and Y. Ohta.
\newblock On the autonomous limit of discrete Painlev{\'e} equations.
\newblock {\em Phys. A} 305 (3):437--444, 2002.
  

\bibitem{Gramani:coalescences}
A.~Ramani and B.~Grammaticos.
\newblock Discrete {P}ainlev\'e equations: coalescences, limits and
  degeneracies.
\newblock {\em Phys. A}, 228(1-4):160--171, 1996.

\bibitem{DPS}
A.~Ramani, B.~Grammaticos and J.~Hietarinta.
\newblock Discrete Versions of the Painlev\'e Equations.
\newblock {\em Phys. Rev. Lett.},  67(14):1829--1832, 1991.

\bibitem{Quadratic}
A.~Ramani, B.~Grammaticos and T. Tamizhmani. 
\newblock Quadratic relations in continuous and discrete Painlev\'e equations. 
\newblock {\em J. Phys. A}, 33(15), 3033, 2000.

\bibitem{Sakai:rational}
H. Sakai.
\newblock Rational surfaces associated with affine root systems and geometry of
  the {P}ainlev\'e equations.
\newblock {\em Comm. Math. Phys.}, 220(1):165--229, 2001.

\bibitem{S88}
E.~K. Sklyanin
\newblock Boundary conditions for integrable quantum systems.
\newblock {\em J. Phys. A: Math. Gen.}, 21:2375-2389, 1988.

\bibitem{TKQ09}
D.~T. Tran, P.~H. van~der~Kamp, and G.~R.~W. Quispel.
\newblock Closed-form expressions for integrals of traveling wave reductions of integrable lattice equations.
\newblock {\em J. Phys. A: Math. Theor.}, 42:225201 (20pp), 2009.

\bibitem{K09}
P.~H. van~der~Kamp.
\newblock Initial value problems for lattice equations.
\newblock {\em J. Phys. A: Math. Theor.}, 42:404019 (16pp), 2009.

\bibitem{KRQ07}
P.~H. van der Kamp, O. Rojas, and G.~R.~W. Quispel.
\newblock Closed-form expressions for integrals of mKdV and sine-Gordon maps.
\newblock {\em J. Phys A.: Math Gen.}, 40:12789--12798, 2007.

\bibitem{KQ10}
P.~H. van~der~Kamp and G.~R.~W. Quispel.
\newblock The staircase method: integrals for periodic reductions of integrable lattice equations.
\newblock {\em J. Phys. A: Math. Theor.}, 43:465207 (34pp), 2010.

\bibitem{Yamada:LaxqEs}
Y. Yamada.
\newblock Lax formalism for {$q$}-{P}ainlev\'e equations with affine {W}eyl
 group symmetry of type {$E^{(1)}_n$}.
\newblock {\em Int. Math. Res. Not. IMRN}, (17):3823--3838, 2011.


\end{thebibliography}

\def\cprime{$'$}

\end{document}